\documentclass[%
 reprint,
 amsmath,amssymb,
 aps,
prb,
]{revtex4-2}

\usepackage{graphicx}% Include figure files
\usepackage{dcolumn}% Align table columns on decimal point
\usepackage{bm}% bold math
\begin{document}

%\preprint{APS/123-QED}

\title{Image of the phonon spectrum in the 1/f noise of topological insulators}% Force line breaks with \\

\author{M. Mihaila}
\email{mihai.mihaila@imt.ro}
\author{S. Dinulescu}%
\author{P. Varasteanu}

 \affiliation{National Institute of Research and Development in Microtechnologies - IMT Bucharest,
Erou Iancu Nicolae str. 126A, 077190, Bucharest, Romania
}

%\affiliation{National Institute of Research and Development in Microtechnologies %- IMT Bucharest,
%Erou Iancu Nicolae str. 126A, 077190, Bucharest, Romania}
 %This line break forced with %\textbackslash\textbackslash
%}%


% \homepage{http://www.Second.institution.edu/~%Charlie.Author}
%\affiliation{
%\affiliation{
% Third institution, the second for Charlie %Author
%\affiliation{%
% This line break forced with %\textbackslash\textbackslash
%}%

%\collaboration{CLEO %Collaboration}%\noaffiliation


\begin{abstract}
%An article usually includes an abstract, a concise summary of the work
%covered at length in the main body of the article. 
%\begin{description}
%use the optional argument of the \verb+\item+ command to give the category of %each item. 
%\end{description}

As reported recently, the 1/f noise intensity in both (Bi,Sb)\textsubscript{2}Te\textsubscript{3}  [Islam \textit{et al.}, Appl. Phys. Lett. \textbf{111}, 062107 (2017)] and BiSbTeSe\textsubscript{1.6} [Biswas \textit{et al.}, Appl. Phys. Lett. \textbf{115}, 131601 (2019)] features noise peaks which develop at some specific temperatures. We compared this noise structure with either phonon density of states or Raman spectrum of each topological insulator (TI), respectively. In (BiSb)\textsubscript{2}Te\textsubscript{3}, the comparison revealed that the noise peaks track the van Hove singularities in the phonon density of states. It resulted that bulk atomic oscillators are responsible for the noise peaks. The most intense noise peak observed in (BiSb)\textsubscript{2}Te\textsubscript{3} at 50 K is attributed to the thermal motion of the Bi atoms. Other less intense noise peaks are assigned to either a single phonon mode or multi-phonon combinations. We found that thermal motions of Bi and Te\textsuperscript{2} atoms in different symmetry directions are involved in most of the phonon combinations, which stand for the signature of the lattice anharmonicity. The noise increase observed in both (Bi,Sb)\textsubscript{2}Te\textsubscript{3} and BiSbTeSe\textsubscript{1.6} above a specific temperature threshold is attributed to the anharmonicity-induced strengthening of the carrier-phonon coupling. In the case of BiSbTeSe\textsubscript{1.6}, we show that all noise singularities are mirrored in the Raman spectrum of a structurally close TI (BiSbTeSe\textsubscript{2}) in the whole temperature range. This indicates that although transport can be at the surface or in the bulk or both of them, the carrier-phonon interaction is the only source of 1/f fluctuations in TIs. Inherently, these results imply that microscopic origin of 1/f noise in solid is in the perpetual thermal motion of the atoms.   
\end{abstract}

\maketitle

%\tableofcontents

For almost a century\cite{Johnson1925}, 1/f noise is puzzling the scientific community with its omnipresence in solid and solid-state devices\cite{Dutta1981,Weissman1988,Hooge1981,Ziel1988,Balandin2013,Fleetwood2020}. In the last decade, it was observed\cite{Hossain2011,Hossain2011a,Cascales2015,Bhattacharyya2015,Bhattacharyya2016,Zhang2016,Islam2017,Biswas2017,Islam2019,Biswas2019,Kunakova2019,Islam2020} in topological insulators (TI), which are quantum materials with metallic conductivity at the surface, while their bulk is insulating\cite{Kane2005,Bernevig2006,Konig2007,Fu2007,Zhang2009,Moore2010,Hasan2010,Ando2013}. These properties are due to the presence of surface electronic states with a Dirac cone energy dispersion located in the bulk band gap\cite{Kane2005,Bernevig2006,Konig2007,Fu2007,Zhang2009,Moore2010,Hasan2010,Ando2013}. 
These surface states are protected by inversion symmetry to backscattering, a property which opens fascinating possibilities for TIs applications in spintronics and quantum topological computing. However, the presence of the 1/f noise in the surface conduction is a factor which can lead to certain technology roadblocks in applications such as quantum computing, where it acts as source of decoherence\cite{Paladino2014,Arute2019}. Therefore, finding its origin is of great practical interest. Hossain \textit{et al.}\cite{Hossain2011a} suggested that the noise dominated by the surface states in TIs could be very low because the time-reversal symmetry topologically protects the surface states from backscattering. However, in Bi\textsubscript{2}Se\textsubscript{3} samples wherein the carrier transport was dominated by the surface, they found a high 1/f noise parameter ($\alpha\approx 0.2$), therefore the noise would be due to a “mixed volume-surface transport regime”\cite{Hossain2011a}. Consequently, to reduce the noise level, it would be necessary to protect the surface conduction from the bulk influences. Nevertheless, in strong quaternary Bi\textsubscript{1.6}Sb\textsubscript{0.4}Te\textsubscript{2}Se, with predominant surface conduction, Bhattacharyya \textit{et al.}\cite{Bhattacharyya2015} found unexpected, bulk-induced “sharp noise peaks at some characteristic”\cite{Bhattacharyya2015} temperatures, while a similar structure was reported by both Islam \textit{et al.}\cite{Islam2017} in (Bi,Sb)\textsubscript{2}Te\textsubscript{3} and Biswas \textit{et al.}\cite{Biswas2019} in BiSbTeSe\textsubscript{1.6} (BSTS). Peaks in the 1/f noise were also reported in tunnel junctions having Bi\textsubscript{2}Te\textsubscript{3} or Bi\textsubscript{2}Se\textsubscript{3} as bottom electrode\cite{Cascales2015}. This situation is similar to the one reported long ago for GaAs Schottky tunnel junction, wherein sharp 1/f noise peaks occurred at voltages corresponding to the GaAs bulk phonon energies\cite{Carruthers1971}. In fact, such a structure in the 1/f noise intensity or $\alpha$ is quite common, for it has been so far observed in many other solid-state physical systems, such as bipolar\cite{Mihaila1984} and MOS transistors\cite{Mihaila2004}, metallic point contacts\cite{Yanson1982,Akimenko1984} , metal films\cite{Mihaila1985, Mihaila2000}, quartz crystals\cite{Planat1987}, carbon nanotubes\cite{Back2009}, carbon soot \cite{Mihaila_patent, Mihaila2019}, metal nanowires \cite{Mihaila2011} or, as predicted \cite{Mihaila_patent}, single molecule\cite{Tsutsui2010,Kim2021}. As for its microscopic origin, the structure in $\alpha$ was found to mirror the van Hove singularities in either phonon density of states (PDOS), $\alpha\approx F(\omega)$, or the Eliashberg function: $\alpha\approx g^{2}(\omega)F(\omega)$, where $g^{2}(\omega)$ is the electron-phonon matrix element and $\omega$ is the phonon frequency\cite{Mihaila2000,Mihaila2019,Mihaila2011,Mihaila1999}. In this work, we show that the noise structure observed by Islam et al.\cite{Islam2017} in (Bi,Sb)\textsubscript{2}Te\textsubscript{3} and Biswas \textit{et al.}\cite{Biswas2019} in BiSbTeSe\textsubscript{1.6} is the image of the phonon spectrum of these materials. To this goal, the noise structure will be compared either with the (Bi,Sb)\textsubscript{2}Te\textsubscript{3} PDOS determined by inelastic neutron scattering (INS)\cite{Klobes2015, Xie2010} or with the Raman spectrum of BiSbTeSe\textsubscript{2}\cite{German2019}. This comparison evidences that each noise peak is associated with a specific phonon mode or multiphonon combinations. The role of lattice anharmonicity in enhancing the noise is brought to the fore. Finally, the significance of these results on the microscopic origin of 1/f noise in solid is briefly discussed.

Figure 1 shows the dependence of the normalized noise intensity, S\textsubscript{V}/V\textsuperscript{2}, on temperature (T) for a 10 nm thick (Bi,Sb)\textsubscript{2}Te\textsubscript{3} film, at f = 1 Hz, in a limited temperature range, as reported by Islam \textit{et al.}\cite{Islam2017}. For comparison purposes, the temperature scale was converted in energy (k\textsubscript{B}T, k\textsubscript{B} - Boltzmann constant).
But, from Hooge’s\cite{Hooge1969} formula: $S_{V}/V^{2} = \alpha/fN$, where $S_{V}$ is the noise spectral density of the fluctuating voltage ($V$) across the sample, at the frequency ($f$) and $N$ is the total charge carrier number in the sample. 
Hence, for f = 1 Hz, one gets $S_{V}/V^{2} = \alpha/N\approx F(\omega)$\cite{Mihaila2000,Mihaila2019,Mihaila1999}. A strong noise peak was observed\cite{Islam2017} at about 50 K (4.3 meV). Other less intense peaks are visible at 25 K (2.15 meV), 82 K (7.06 meV), 110 K (9.47 meV), 140 K (12.06 meV) and 178 K (15.33 meV).
\begin{figure}
\includegraphics[scale = 0.3]{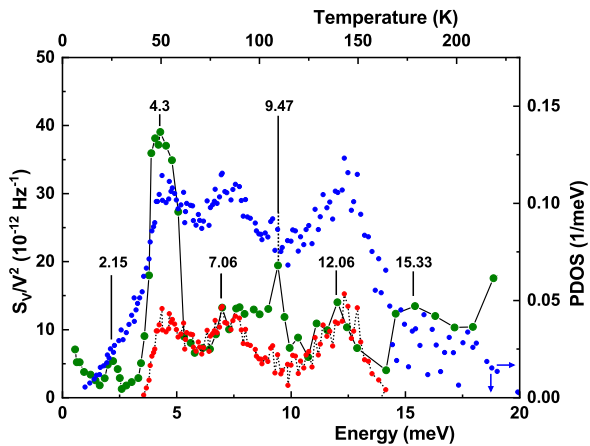}
\caption{\label{fig:fig1} Comparison between the temperature dependence of $S_{V}/V^{2}$ in (Bi,Sb)\textsubscript{2}Te\textsubscript{3} (green dots - reprinted from Ref. \cite{Islam2017}, S. Islam \textit{et al.}, Appl. Phys. Lett. \textbf{111}, 062107 (2017), with permission of AIP Publishing) and PDOS of (Bi\textsubscript{0.26}Sb\textsubscript{0.74})\textsubscript(2)Te\textsubscript{3} (blue and red dots, - reproduced (adapted) with permission from Ref. \cite{Klobes2015}- B. Klobes \textit{et al.} Phys. Stat. Sol. 9, 57, 2015; copyright 2015 John Wiley and Sons); the numbers denote noise peak energies; dotted line is guide to the eye.}
\end{figure}

In a preliminary investigation, the noise peak energies were compared with the phonon energies of both Bi\textsubscript{2}Te\textsubscript{3} and Sb\textsubscript{2}Te\textsubscript{3} binary compounds. It was found that the noise peak at 25 K (2.15 meV) corresponds to a 0.51 THz (2.11 meV) transverse acoustic phonon ($E_{u}^{1}$) in Bi\textsubscript{2}Te\textsubscript{3}, measured by INS at 77 K, at the Z point of symmetry\cite{Wagner1978,Kullmann1990}. At 25 K, its frequency, which decreases with temperature\cite{Wagner1978} with the slope $-3.139\times10^{-4}$ THz/K (linear approximation), becomes 0.526 THz, namely 2.17 meV, an energy which reasonably fits the noise one. Surprisingly, the frequency of this ($E_{u}^{1}$) phonon at 50 K is 0.518 THz (2.145 meV), so that, fortuitous or not, the most intense noise peak at 4.3 meV seems to be associated with the energy (4.29 mV) of its second ($2\times2.145$ meV) overtone. Also, the average energy (4.3 meV) of the $E_{g}^{1}$ Raman modes in Bi\textsubscript{2}Te\textsubscript{3} and Sb\textsubscript{2}Te\textsubscript{3} layers, observed at 4.26 meV (34.4 cm\textsuperscript{-1})\cite{Shahil2010} and 4.35 meV (35.1 cm\textsuperscript{-1})\cite{Shahil2012}, respectively, is in agreement with the noise data. The small noise peak at 7.06 meV (82 K) is close to the 6.95 meV (56 cm\textsuperscript{-1}, $E_{u}^{1}$(TO) infrared phonon observed in Sb\textsubscript{2}Te\textsubscript{3} at 80 K\cite{Richter1982}.The noise peak at 9.47 meV (110 K) is close to the $E_{u}^{1}$ transversal mode at 2.25 THz (9.31 meV) at 77 K, measured in Bi\textsubscript{2}Te\textsubscript{3} at the Z point of symmetry \cite{Wagner1978}.  At 110 K, its frequency, which increases with temperature with the slope $7.623\times10^{-4}$THz/K \cite{Wagner1978}, is 2.275 THz (9.42 meV), a value consistent with 9.47 meV from noise measurement. Also, a 9.49 meV (76.5 cm\textsuperscript{-1}) phonon mode was reported\cite{Richter1982} at 80 K in Sb\textsubscript{2}Te\textsubscript{2} for the $E_{u}^{1}$(LO) infrared mode. A sharp $E_{u}^{2}$ phonon peak was found in the Eliashberg function of Bi\textsubscript{2}Te\textsubscript{3} by Inelastic Electron Tunneling Spectroscopy(IETS)\cite{Nagao1998} at 12 meV. Although measured at 4.2 K, it is strikingly close to the noise peak at 12.06 meV (140 K). However, from Richter \textit{et al.}\cite{Richter1977}, the $E_{u}^{2}$ phonon frequency at 140 K in Bi\textsubscript{2}Te\textsubscript{3}, which shifts with -0.0105 cm\textsuperscript{-1}/K, is 96.69 cm\textsuperscript{-1} or 11.99 meV, a value close to 12.06 meV from noise data. As for the enlarged noise band with the maximum at 15.33 meV (178 K), it corresponds to the 3.7 THz (15.32 meV) $A_{2u}^{2}$ phonon at the $\Gamma$ point, measured\cite{Wagner1978} in Bi\textsubscript{2}Te\textsubscript{3} at 77 K or to 15.4 meV infrared $E_{u}^{2}$ mode in Sb\textsubscript{2}Te\textsubscript{3} at 80 K\cite{Richter1982}. All these comparisons are synthetically presented in the Table I.

The foregoing analysis showed that the temperatures where the noise peaks develop are closely related to the phonon energies of the binary compounds. The noise data were further compared with PDOS of Bi\textsubscript{2}Te\textsubscript{3} \cite{Rauh1981}, Sb\textsubscript{2}Te\textsubscript{3}\cite{Rauh1981}, (Bi\textsubscript{0.26}Sb\textsubscript{0.74})\textsubscript{2}Te\textsubscript{3}\cite{Klobes2015} and (Bi,Sb)\textsubscript{2}Te\textsubscript{3}\cite{Xie2010} compounds, all obtained by INS. The comparison between the noise structure and the bulk PDOS of the as-cast (Bi\textsubscript{0.26}Sb\textsubscript{0.74})\textsubscript{2}Te\textsubscript{3}, obtained by Klobes \textit{et al.} \cite{Klobes2015} at 295 K, is shown in Figure.~\ref{fig:fig1}. According to these authors \cite{Klobes2015}, the phonon spectrum of this material is similar to (Bi,Sb)\textsubscript{2}Te\textsubscript{3}. The comparison revealed that the first noise peak at 2.15 meV corresponds to a shoulder in PDOS. Strikingly, the dominant noise peak at 4.3 meV fits the first PDOS peak. Next, the PDOS part located in the (3.5-14.2) meV energy range was downward shifted and scaled down till its second peak fitted the intensity of the 7.06 meV noise peak (red dots in Figure.~\ref{fig:fig1}). This procedure unraveled a good fit between the noise peaks and PDOS structure. It also revealed that the third major PDOS peak is well mirrored in the noise peak at 12.06 meV. The strongly scattered PDOS points in the (14-18) meV energy range are surprisingly well enveloped by the large noise band with the maximum at 15.33 meV. Although the second as intensity, the noise peak at 9.47 meV corresponds to a clear but less intense PDOS peak. Figure.~\ref{fig:fig1} shows that the dominant PDOS peaks feature almost the same intensity, while the intensity of their noise counterparts is quite different. This indicates that besides $F(\omega)$, there is another factor involved in the noise intensity control, which, as stated previously\cite{Mihaila2000,Mihaila2019,Mihaila2011,Mihaila1999}, could be the carrier-phonon coupling matrix element. This interaction should depend on the nature of the atoms the carriers interact with. Since at low energies the phonon spectrum is dominated by thermal vibration of the heaviest element (Bi)\cite{Rauh1981}, one would expect that the noise peaks at lower energies to be related to the group V element (Bi), while those at higher energies can be due to the modes associated with the group VI element (Te)\cite{Rauh1981}. This statement is supported by the existence of an intense peak around 4.3 meV in the Bi partial density of states in Bi\textsubscript{2}Te\textsubscript{3} \cite{Klobes2015} Also, using neutron diffraction, Serrano-Sanchez \textit{et al.} \cite{SerranoSanchez2017} found that in-plane motion of the Bi atoms in the direction [$1\overline{1}0$], which is the major, most favored axis of their anisotropic displacements in Bi\textsubscript{0.35}Sb\textsubscript{1.65}Te\textsubscript{3}, corresponds to a 4.4 meV Einstein oscillator. On these reasons, we tentatively attribute the strongest noise peak at 4.3 meV to the interaction of the carriers with the thermal motion of the Bi atoms (Table I). Since the Bi atoms move in a strong anharmonic potential \cite{Wagner1978, Kullmann1990, Cheng2011, Hellman2014}, a stronger carrier-phonon coupling is expected, therefore more noise \cite{Mihaila2000,Mihaila2019,Mihaila1999}. A 6.9 meV Einstein oscillator has been reported \cite{SerranoSanchez2017} for the out-of-plane motion of the Te\textsuperscript{2} atoms in the direction [001], which is close to the 7.06 meV noise peak energy.  Using the same data \cite{SerranoSanchez2017}, values of 9.5 meV or 9.4 meV, resulting from combinations 2$\times$Te\textsuperscript{1}[001]-Bi[001] and 3$\times$Te\textsuperscript{2}[$1\overline{1}0$]-Bi[$1\overline{1}0$], respectively, are close to the 9.47 meV noise peak energy, while the 12.06 meV noise peak corresponds to a combination 4.6 + 7.4=12 meV of the Te\textsuperscript{2}+Te\textsuperscript{1} motion in the directions [$1\overline{1}0$] and [001], respectively. Although the 15.33 meV noise band is located in a PDOS region dominated by both Sb and Te thermal motion \cite{Klobes2015, Rauh1981}, it also fits the combination 3$\times$7.4-6.9 = 15.3 meV, where 7.4 meV and 6.9 meV are the energies of the Te\textsuperscript{1} and Te\textsuperscript{2} out-of plane atomic motion ([001], parallel to the c-axis), respectively\cite{SerranoSanchez2017}. All these data complete the Table I.

\begin{table*}
\caption{\label{tab:table1}Comparison between noise peak energies in (BiSb)\textsubscript{2}Te\textsubscript{3}\cite{Islam2017} and phonon energies in Bi\textsubscript{2}Te\textsubscript{3}, Sb\textsubscript{2}Te\textsubscript{3} and Bi\textsubscript{0.35}Sb\textsubscript{1.65}Te\textsubscript{3} determined by INS, Raman (R), infrared (IR) and neutron diffraction (ND) spectroscopy. Te\textsuperscript{1} and Te\textsuperscript{2} denote the two Te atoms in the Te\textsuperscript{1}-Bi-Te\textsuperscript{2}-Bi-Te\textsuperscript{2} quintuple layer.}
%\begin{ruledtabular}
\begin{tabular}{|c|c|c|c|c|c|} \hline
 \textbf{Peak}&\textbf{T (K)}&\textbf{$\bm{k_{B}T}$ from noise (meV)}&\textbf{Phonon energy (meV)}
&\textbf{Phonon mode or combinations, TI}&\textbf{Ref./Method}\\ \hline
 1&25 &2.15 &2.17&$E_{u}^{1}$(25 K),Bi\textsubscript{2}Te\textsubscript{2}&\cite{Wagner1978,Kullmann1990}/INS\\ \hline
 2&50
 &4.3&\begin{tabular}[c]{@{}c@{}}$4.29 = 2\times2.145$\\ $4.3 = (4.26 + 4.35)/2$\\ $4.4$\end{tabular}&\begin{tabular}[c]{@{}c@{}}$2\times E_{u}^1$ (50 K), Bi\textsubscript{2}Te\textsubscript{3}\\ $E_{g}^{1}$, (Bi\textsubscript{2}Te\textsuperscript{3} + Sb\textsubscript{2}Te\textsubscript{3})/2\\ Bi$[1\overline{1}0]$, Bi\textsubscript{0.35}Sb\textsubscript{1.65}Te\textsubscript{3}\end{tabular}&\begin{tabular}[c]{@{}c@{}}\cite{Wagner1978,Kullmann1990}/INS\\ \cite{Shahil2010,Shahil2012}/R\\ \cite{SerranoSanchez2017}/ND\end{tabular}\\ \hline
3&82&7.06&\begin{tabular}[c]{@{}c@{}}6.95\\ 6.9\end{tabular}&\begin{tabular}[c]{@{}c@{}}$E_{u}^{1}$(TO) (80 K), Sb\textsubscript{2}Te\textsubscript{3}\\ Te\textsuperscript{2}[001],Bi\textsubscript{0.35}Sb\textsubscript{1.65}Te\textsubscript{3}\end{tabular}&\begin{tabular}[c]{@{}c@{}}\cite{Richter1982}/IR\\\cite{SerranoSanchez2017}/ND\end{tabular}\\ \hline

4&110&9.47&\begin{tabular}[c]{@{}c@{}}9.42\\9.49\\ $9.5 = 2\times 7.4 - 5.3$ \\ $9.4 = 3\times 4.6 - 4.4$\end{tabular}&
\begin{tabular}[c]{@{}c@{}}$E_{u}^{1}$(110 K), Bi\textsubscript{2}Te\textsubscript{3}\\$E_{u}^{1}$(LO)(80 K), Sb\textsubscript{2}Te\textsubscript{3}\\ $2\times $Te\textsuperscript{1}[001] - Bi[001], Bi\textsubscript{0.35}Sb\textsubscript{1.65}Te\textsubscript{3} \\ $3\times$ Te\textsuperscript{2}$[1\overline{1}0]$ - Bi$[1\overline{1}0]$, Bi\textsubscript{0.35}Sb\textsubscript{1.65}Te\textsubscript{3}\end{tabular}&\begin{tabular}[c]{@{}c@{}}\cite{Wagner1978}/INS\\ \cite{Richter1982}/IR\\ \cite{SerranoSanchez2017}/ND\\\cite{SerranoSanchez2017}/ND\end{tabular}\\ \hline

5&140&12.06&\begin{tabular}[c]{@{}c@{}}12\\11.99\\ 12 = 4.6 + 7.4\end{tabular}& \begin{tabular}[c]{@{}c@{}}$A_{2u}^{2}$ (77 K), Bi\textsubscript{2}Te\textsubscript{3}\\ $E_{u}^{2}$ (80 K), Sb\textsubscript{2}Te\textsubscript{3}\\$3\times $Te\textsuperscript{1}$[1\overline{1}0]$ + Te\textsuperscript{2}[001], Bi\textsubscript{0.35}Sb\textsubscript{1.65}Te\textsubscript{3}\end{tabular}&\begin{tabular}[c]{@{}c@{}}\cite{Nagao1998}/IETS\\ \cite{Richter1977}/IR\\ \cite{SerranoSanchez2017}/ND\end{tabular} \\ \hline

6&178&15.33&\begin{tabular}[c]{@{}c@{}}15.32\\ 15.4\\ 15.3 = $3\times 7.4-6.9$\end{tabular}&
\begin{tabular}[c]{@{}c@{}}$A_{2u}^{2}$ (77 K), Bi\textsubscript{2}Te\textsubscript{3}\\ $E_{u}^{2}$ (80 K), Sb\textsubscript{2}Te\textsubscript{2}\\ $3\times $Te\textsuperscript{1}[001] - Te\textsuperscript{2}[001], Bi\textsubscript{0.35}Sb\textsubscript{1.65}Te\textsubscript{3}\end{tabular} &
\begin{tabular}[c]{@{}c@{}}\cite{Wagner1978}/INS\\ \cite{Richter1982}/IR\\ \cite{SerranoSanchez2017}/ND\end{tabular} \\ \hline

7&240&20.68&\begin{tabular}[c]{@{}c@{}}20.7 = $3\times 6.9$\\ 20.7 = $3\times 4.6 + 6. 9$\\ $20.7 = 4.9 + 13.8 $\end{tabular} & \begin{tabular}[c]{@{}c@{}}$3\times $Te\textsuperscript{2}[001], Bi\textsubscript{0.35}Sb\textsubscript{1.65}Te\textsubscript{3}\\ $3\times $ Te\textsuperscript{2}$[1\overline{1}0]$ + Te\textsuperscript{2}[110], Bi\textsubscript{0.35}Sb\textsubscript{1.65}Te\textsubscript{3}\\ Te\textsuperscript{2}[001] + Te\textsuperscript{2}[110], Bi\textsubscript{0.35}Sb\textsubscript{1.65}Te\textsubscript{3}\end{tabular}&
\begin{tabular}[c]{@{}c@{}}\cite{SerranoSanchez2017}/ND\\ \cite{SerranoSanchez2017}/ND\\ \cite{SerranoSanchez2017}/ND \end{tabular} \\ \hline

8&290&24.98&\begin{tabular}[c]{@{}c@{}}$24.9 = 2\times 5.3 + 7.4 + 6.9$\\ $24.9 = 4.4 + 6.7 + 2\times 6.9$\\ $24.9 = 4.4 + 6.7 + 3\times 4.6$ \\ \\ \end{tabular} & \begin{tabular}[c]{@{}c@{}}$2\times$Bi[001] + Te\textsuperscript{1}[001] + Te\textsuperscript{2}[001]\\ Bi$[1\overline{1}0]$ + Te\textsuperscript{1}$[1\overline{1}0] + 2\times$Te\textsuperscript{2}[001] \\ Bi$[1\overline{1}0]$ + Te\textsuperscript{1}$[1\overline{1}0] + 3\times $Te\textsuperscript{2}$[1\overline{1}0]$ \\ all in Bi\textsubscript{0.35}Sb\textsubscript{1.65}Te\textsubscript{3}\end{tabular} & 
\begin{tabular}[c]{@{}c@{}}\cite{SerranoSanchez2017}/ND\\ \cite{SerranoSanchez2017}/ND\\ \cite{SerranoSanchez2017}/ND \\ \\ \end{tabular} \\

\hline

\end{tabular}
%\end{ruledtabular}
\end{table*}

\begin{figure}
\includegraphics[scale = 0.3]{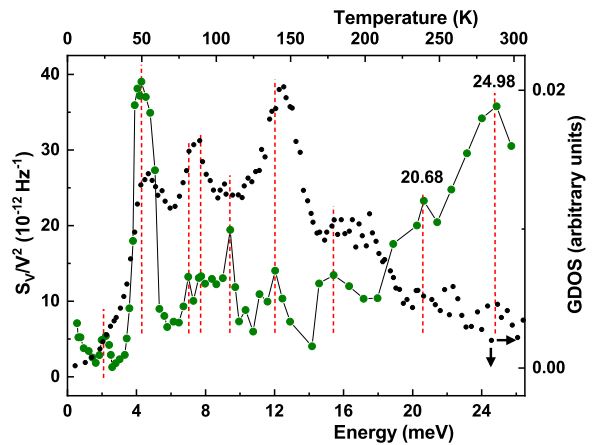}% Here is how to import EPS art
\caption{\label{fig:fig2} Comparison between temperature dependence of $S_{V}/V^{2}$ in (Bi,Sb)\textsubscript{2}Te\textsubscript{3} (green dots - reprinted from Ref. \cite{Islam2017}, S. Islam \textit{et al.}, Appl. Phys. Lett. \textbf{111}, 062107 (2017), with permission of AIP Publishing) and (Bi,Sb)\textsubscript{2}Te\textsubscript{3} (P/G)DOS (black dots, reprinted/adapted with permission from Ref. \cite{Xie2010} - W. Xie \textit{et al.}, Nano Lett. 10, 3283 (2010); copyright (2010) American Chemical Society); red, dashed lines indicate the correspondence noise structure - phonon spectrum, while numbers denote noise peak energies.}
\end{figure}

\begin{figure}
\includegraphics[scale = 0.3]{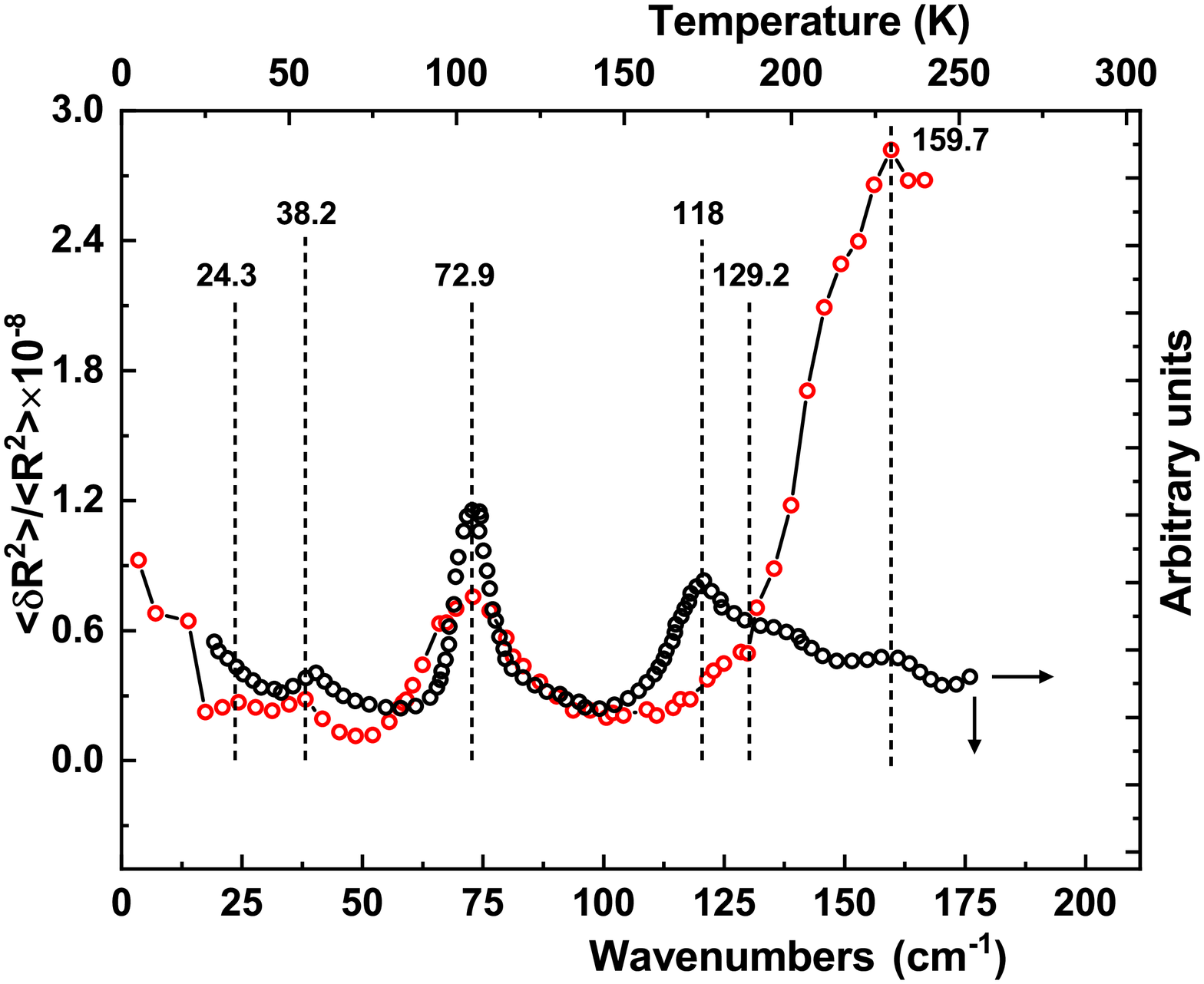}% Here is how to import EPS art
\caption{\label{fig:fig3} Comparison between the temperature dependence of $<$$\delta R^{2}$$>$/$<$$R^{2}$$>$ (red circles) in BiSbTeSe\textsubscript{1.6} (Reprinted from Ref. \cite{Biswas2019}, Biswas \textit{et al.}, Appl. Phys. Lett. \textbf{115}, 131601 (2019), with permission of AIP Publishing) and Raman spectrum of BiSbTeSe\textsubscript{2} (black circles, reprinted with permission from Ref. \cite{German2019} - R. German \textit{et al.}, Phys. Rev. Mat. \textbf{3}, 054204 (2019); copyright 2019 by the American Physical Society); the correspondence noise structure-Raman spectrum is indicated by discontinuous lines, while numbers denote the noise peak energies in equivalent wavenumbers (cm\textsuperscript{-1}).}
\end{figure}

The role of the electron-phonon coupling in the mechanism of noise appears even more pregnant from Figure.~\ref{fig:fig2}, where the complete noise data of Islam \textit{et. al.} \cite{Islam2017} are compared with the (Bi,Sb)\textsubscript{2}Te\textsubscript{3} phonon spectrum obtained by Xie \textit{et al.}\cite{Xie2010} by INS, at 300 K. In this case, a small difference, due to anharmonicity, between the noise peaks and the phonon ones exists. One also notes that the intensity of the dominant peaks in the phonon spectrum is quite different. It can be seen that while the phonon density increases, the noise intensity decreases and vice-versa for temperature higher than about 200K: the PDOS slowly decreases, while the noise intensity increases. In this temperature range, two new noise peaks develop at 240 K (20.68 meV) and 290 K (24.98 meV) (Table I), which reasonably fit some bulges in the phonon spectrum.
In terms of Einstein oscillators \cite{SerranoSanchez2017}, both the 3$\times$Te\textsuperscript{2}[001] overtone (3$\times$6.9=20.7 meV) and combinations such as 3$\times$Te\textsuperscript{2}[$1\overline{1}0$]+Te\textsuperscript{2}[001] or Te\textsuperscript{2}[001]+Te\textsuperscript{2}[110] gave the same result (20.7 meV, Table I), close to the noise peak at 20.68 meV. The noise peak at 24.98 meV (290 K) is well fitted by 24.9 meV, which results from the four-phonon 2$\times$Bi[001]+Te\textsuperscript{2}[001] + Te\textsuperscript{1}[001], Bi[$1\overline{1}0$]+Te1[$1\overline{1}0$]+2$\times$Te\textsuperscript{2}[001] or Bi[$1\overline{1}0$]+3$\times$Te\textsuperscript{2}[$1\overline{1}0$]+Te\textsuperscript{1}[$1\overline{1}0$]\cite{SerranoSanchez2017} five-phonon combinations (Table I). Such multi-phonon combinations are the signatures of lattice anharmonicity.
One observes that single phonons and their overtones associated with the thermal motion of Bi and, especially, Te\textsuperscript{2} atoms are involved in most of these multi-phonon combinations. This seems to be the manifestation of lattice anharmonicity, because, as shown by Serrano-Sanchez \textit{et al.}\cite{SerranoSanchez2017}, Bi and Te\textsuperscript{2} are the most anharmonic oscillators in Bi\textsubscript{0.35}Sb\textsubscript{1.65}Te\textsubscript{3}. One can thus attribute the noise increase starting around 200 K to an enhanced carrier-phonon coupling induced by the stronger anharmonicity installed in the sample at this temperature. A similar noise increase was observed in Bi\textsubscript{1.6}Sb\textsubscript{0.4}Te\textsubscript{2}Se at the same temperature \cite{Bhattacharyya2015}, while in BiSbTeSe\textsubscript{1.6} Biswas \textit{et al.} \cite{Biswas2019} showed that the noise starts increasing slowly at about 160 K \cite{Biswas2017} (Figure.~\ref{fig:fig3}), followed by a sudden increase at a kink located around 186K and culminated with a noise peak at 230 K. 
The relative resistance (R) noise of Biswas \textit{et al.}\cite{Biswas2019} is compared in Figure.~\ref{fig:fig3} with the Raman spectrum of BiSbTeSe\textsubscript{2} collected by German \textit{et al.} \cite{German2019} in parallel polarization, c(aa)c geometry, at 5 K. The equivalent wavenumber (24.3 cm\textsuperscript{-1}, 1 K = 0.6945 cm\textsuperscript{-1}) of the first noise peak at 35 K (U\textsubscript{n} in the Table II) has no correspondent in the Raman spectrum of Ref. \cite{German2019}. The noise peak at 55 K (38.2 cm\textsuperscript{-1}) is associated with the E\textsubscript{g} phonon \cite{German2019} which is 39.8 cm\textsuperscript{-1} at 55 K. The prominent noise peak at 105 K (72.92 cm\textsuperscript{-1}) is well mirrored in the Raman spectrum by a 72.7 cm\textsuperscript{-1} peak at 5 K \cite{German2019}, which shifts to about 71.5 cm\textsuperscript{-1} at 105 K (from Figure. 3a of Ref. \cite{German2019}). A weak but clear shoulder in the noise data at about 170 K (118 cm\textsuperscript{-1}) corresponds to the 5\textsuperscript{th} phonon mode of German \textit{et al.}\cite{German2019}, with a frequency of 118 cm\textsuperscript{-1} at 170 K (from Figure. 3b of Ref. \cite{German2019}). The noise kink at 186 K(129.2 cm\textsuperscript{-1}) corresponds to the 6\textsuperscript{th} BSTS Raman phonon of German \textit{et al.}\cite{German2019}, for its frequency at 186 K  (slope -2.56$\times$10-2 cm\textsuperscript{-1}/K)  is 129.5 cm\textsuperscript{-1}. The strongest noise peak at 230 K (159.7 cm\textsuperscript{-1}) is close to 160.2 cm\textsuperscript{-1}, an $A_{1g}$ phonon in BSTS \cite{German2019}, which shows the highest full width at the halve maximum (FWHM) among all Raman peaks of Ref. \cite{German2019}. Since the FWHM is a measure of the carrier-phonon coupling, the enhanced noise can be explained by a stronger coupling. In addition, the anharmonicity involvement in noise generation is supported by the observation that the third noise peak is very well described by the third overtone 3$\times$24.3 = 72.9 cm\textsuperscript{-1} of the first noise peak. Also, the six phonons combination (5$\times$24.3 + 38.2) of the first two noise peak energies gives a similar value (159.7 cm\textsuperscript{-1}) for the 6\textsuperscript{th} noise peak. All these correlations are presented in the Table II.

\begin{table*}
\caption{\label{tab:table2}Comparison between the noise peak energies in BiSbTeSe\textsubscript{1.6} \cite{Biswas2019} and phonon energies of BiSbTeSe\textsubscript{2} determined by Raman spectroscopy}
%\begin{ruledtabular}
\begin{tabular}{|c|c|c|c|c|c|} \hline
 \textbf{Peak}&\textbf{T (K)}&\textbf{$\bm{k_{B}T}$ from noise (cm\textsuperscript{-1})}&\textbf{Phonon from Raman spectroscopy (cm\textsuperscript{-1})}&\textbf{Phonon mode}
&\textbf{Ref.}\\ \hline

1&35&24.3&-&U\textsubscript{n}&- \\
\hline
2&55&38.2&39.8&$E_{g}^{1}$&\cite{German2019} \\
\hline
3&105&72.9&\begin{tabular}[c]{@{}c@{}}71.5 (at 105 K)\\$72.9 = 3\times 24.3$ \end{tabular}&
\begin{tabular}[c]{@{}c@{}}$A_{1g}$\\ 3U\textsubscript{n} \end{tabular}&\begin{tabular}[c]{@{}c@{}}\cite{German2019}\\ \cite{Biswas2019} \end{tabular} \\   \hline

4&170&118&118 (at 170 K)&$E_{g}$&\cite{German2019} \\   \hline

5&186&129.2&129.2 (at 186 K)&$A_{1g}$&\cite{German2019} \\   \hline

6&230&159.7&\begin{tabular}[c]{@{}c@{}}160.2 (294 K)\\$159.7 = 5\times 24.3 + 38.2$ \end{tabular}&
\begin{tabular}[c]{@{}c@{}}$A_{1g}^{2}$\\ 5U\textsubscript{n} + $E_g^1$ \end{tabular}&\begin{tabular}[c]{@{}c@{}}\cite{German2019}\\ \cite{Biswas2019} \end{tabular} \\   \hline

\end{tabular}
%\end{ruledtabular}
\end{table*}

Carrier hopping in the bulk was the transport mechanism found by Biswas et al.\cite{Biswas2019} in BSTS at temperature higher than 150 K. Voss\cite{Voss1978}, who found hopping in a MOSFET inversion layer at 4.2 K, concluded that 1/f noise “is an intrinsic transport property of the hopping conduction”\cite{Voss1978}. As shown elsewhere \cite{Mihaila2004}, Voss’ statement implies phonons contribution because “hopping can only occur if phonons can be emitted or absorbed to compensate for the change in energy of the electron”\cite{Thouless1974}. Therefore, 1/f noise in BSTS sample could be generated by a phonon-assisted hopping mechanism. However, as shown by authors \cite{Biswas2019}, this noise source can be obscured by the presence of the random telegraph signal noise. It is thus not clear whether the noise structure in this temperature domain is exclusively a 1/f noise-related property. However, for T $<$ 150 K, the normalized noise $fS_{R}/R^{2}$ is (almost) flat \cite{Biswas2019}, hence 1/f source is dominant. In this temperature domain, a surface conduction channel is active \cite{Biswas2019}, therefore a surface 1/f noise source may exist, which, according to Biswas et al.\cite{Biswas2019}, is due to “coupling of surface transport with defect dynamics in bulk”\cite{Biswas2019}. Similarly, Islam et al.\cite{Islam2017} concluded that “the internal bulk chalcogenide defects are mainly responsible for resistance fluctuations for both surface and bulk electronic states”\cite{Islam2017}. Since the noise structure correlates with the bulk phonon energies in the whole temperature range, it results that wherever the current flows, the carriers interaction with the bulk phonons is the unique microscopic source of noise in both TIs. For decades, it has been shown that 1/f noise in solid is generated at surface \cite{McWhorter1955,Sah1966,Hsu1968,Koslowski1993,Sugita1996} or in bulk \cite{Hooge1969, Fleetwood1983, Zimmerman1986}, while other works proved that the noise source can be shifted from the surface to the bulk \cite{Liu2013} or vice-versa \cite{Akarvardar2006}.  Phonons appear to be involved in both cases, for surface and bulk phonons have been observed in the noise of the same physical system \cite{Mihaila1991,Mihaila2003}, while specific phonon properties\cite{Asriyan2005} were exploited to control interface noise\cite{Hammig2013}. Also, fluctuations in the number of both surface \cite{Koslowski1993,Sugita1996} and bulk phonons\cite{Toshimitsu1990} were proposed as 1/f noise sources. The analysis of the noise data in TIs revealed a new situation in solid: the fluctuation of the surface conductivity is controlled by the bulk phonons. The association of the noise peaks in TIs with the thermal motion of different atomic oscillators supports the earlier findings which considered that microscopic origin of 1/f noise in solid is in the “perpetual equilibrium atomic motion”\cite{Mihaila2000,Mihaila2011}. This mechanism can explain 1/f noise in any solid-state system, in both equilibrium \cite{Voss1976} and nonequilibrium \cite{Mihaila2019}.

In conclusion, the 1/f noise structure observed by Islam et al.\cite{Islam2017} in (Bi,Sb)\textsubscript{2}Te\textsubscript{3} was correlated with the van Hove singularities in the bulk phonon spectrum of this TI. Thermal motion of the Bi atoms was identified as the source of the most intense noise peak observed in this material at 50 K. Other noise peaks were assigned to multi-phonon combinations, which are the signatures of (Bi,Sb)\textsubscript{2}Te\textsubscript{3} lattice anharmonicity. We found that the 1/f noise structure reported by Biswas et al.\cite{Biswas2019} in BiSbTeSe\textsubscript{1.6} tracks the BiSbTeSe\textsubscript{2} Raman spectrum, each noise singularity being assigned to a single- or multi-phonon energy. Stronger carrier-phonon coupling due to anharmonicity was invoked as explanation of the noise increase observed above some specific temperature thresholds in both materials. The observation of the phonon structure in the whole temperature range strongly supports the carrier-phonon interaction as the unique microscopic source of 1/f noise at the surface or in the bulk. Our investigation revealed that the thermal motion of different bulk atoms is involved in the mechanism of 1/f noise in topological insulators. The noise data in TI supports the thermal motion of the atoms as microscopic source of 1/f noise in solid.


%\bibliography{LATEX_final_corectat_Mendeley}% Produces the bibliography via BibTeX.

\begin{thebibliography}{76}%
\makeatletter
\providecommand \@ifxundefined [1]{%
 \@ifx{#1\undefined}
}%
\providecommand \@ifnum [1]{%
 \ifnum #1\expandafter \@firstoftwo
 \else \expandafter \@secondoftwo
 \fi
}%
\providecommand \@ifx [1]{%
 \ifx #1\expandafter \@firstoftwo
 \else \expandafter \@secondoftwo
 \fi
}%
\providecommand \natexlab [1]{#1}%
\providecommand \enquote  [1]{``#1''}%
\providecommand \bibnamefont  [1]{#1}%
\providecommand \bibfnamefont [1]{#1}%
\providecommand \citenamefont [1]{#1}%
\providecommand \href@noop [0]{\@secondoftwo}%
\providecommand \href [0]{\begingroup \@sanitize@url \@href}%
\providecommand \@href[1]{\@@startlink{#1}\@@href}%
\providecommand \@@href[1]{\endgroup#1\@@endlink}%
\providecommand \@sanitize@url [0]{\catcode `\\12\catcode `\$12\catcode
  `\&12\catcode `\#12\catcode `\^12\catcode `\_12\catcode `\%12\relax}%
\providecommand \@@startlink[1]{}%
\providecommand \@@endlink[0]{}%
\providecommand \url  [0]{\begingroup\@sanitize@url \@url }%
\providecommand \@url [1]{\endgroup\@href {#1}{\urlprefix }}%
\providecommand \urlprefix  [0]{URL }%
\providecommand \Eprint [0]{\href }%
\providecommand \doibase [0]{https://doi.org/}%
\providecommand \selectlanguage [0]{\@gobble}%
\providecommand \bibinfo  [0]{\@secondoftwo}%
\providecommand \bibfield  [0]{\@secondoftwo}%
\providecommand \translation [1]{[#1]}%
\providecommand \BibitemOpen [0]{}%
\providecommand \bibitemStop [0]{}%
\providecommand \bibitemNoStop [0]{.\EOS\space}%
\providecommand \EOS [0]{\spacefactor3000\relax}%
\providecommand \BibitemShut  [1]{\csname bibitem#1\endcsname}%
\let\auto@bib@innerbib\@empty
%</preamble>
\bibitem [{\citenamefont {Johnson}(1925)}]{Johnson1925}%
  \BibitemOpen
  \bibfield  {author} {\bibinfo {author} {\bibfnamefont {J.~B.}\ \bibnamefont
  {Johnson}},\ }\bibfield  {title} {\bibinfo {title} {{The Schottky effect in
  low frequency circuits}},\ }\href {https://doi.org/10.1103/PhysRev.26.71}
  {\bibfield  {journal} {\bibinfo  {journal} {Phys. Rev.}\ }\textbf {\bibinfo
  {volume} {26}},\ \bibinfo {pages} {71} (\bibinfo {year} {1925})}\BibitemShut
  {NoStop}%
\bibitem [{\citenamefont {Dutta}\ and\ \citenamefont {Horn}(1981)}]{Dutta1981}%
  \BibitemOpen
  \bibfield  {author} {\bibinfo {author} {\bibfnamefont {P.}~\bibnamefont
  {Dutta}}\ and\ \bibinfo {author} {\bibfnamefont {P.~M.}\ \bibnamefont
  {Horn}},\ }\bibfield  {title} {\bibinfo {title} {{Low-frequency fluctuations
  in solids: 1/f noise}},\ }\href {https://doi.org/10.1103/RevModPhys.53.497}
  {\bibfield  {journal} {\bibinfo  {journal} {Rev. Mod. Phys.}\ }\textbf
  {\bibinfo {volume} {53}},\ \bibinfo {pages} {497} (\bibinfo {year}
  {1981})}\BibitemShut {NoStop}%
\bibitem [{\citenamefont {Weissman}(1988)}]{Weissman1988}%
  \BibitemOpen
  \bibfield  {author} {\bibinfo {author} {\bibfnamefont {M.~B.}\ \bibnamefont
  {Weissman}},\ }\bibfield  {title} {\bibinfo {title} {{1/f Noise and other
  slow, nonexponential kinetics in condensed matter}},\ }\href
  {https://doi.org/10.1103/RevModPhys.60.537} {\bibfield  {journal} {\bibinfo
  {journal} {Rev. Mod. Phys.}\ }\textbf {\bibinfo {volume} {60}},\ \bibinfo
  {pages} {537} (\bibinfo {year} {1988})}\BibitemShut {NoStop}%
\bibitem [{\citenamefont {Hooge}\ \emph {et~al.}(1981)\citenamefont {Hooge},
  \citenamefont {Kleinpenning},\ and\ \citenamefont {Vandamme}}]{Hooge1981}%
  \BibitemOpen
  \bibfield  {author} {\bibinfo {author} {\bibfnamefont {F.~N.}\ \bibnamefont
  {Hooge}}, \bibinfo {author} {\bibfnamefont {T.~G.}\ \bibnamefont
  {Kleinpenning}},\ and\ \bibinfo {author} {\bibfnamefont {L.~K.}\ \bibnamefont
  {Vandamme}},\ }\bibfield  {title} {\bibinfo {title} {{Experimental studies on
  1/f noise}},\ }\href {https://doi.org/10.1088/0034-4885/44/5/001} {\bibfield
  {journal} {\bibinfo  {journal} {Reports Prog. Phys.}\ }\textbf {\bibinfo
  {volume} {44}},\ \bibinfo {pages} {479} (\bibinfo {year} {1981})}\BibitemShut
  {NoStop}%
\bibitem [{\citenamefont {{Van Der Ziel}}(1988)}]{Ziel1988}%
  \BibitemOpen
  \bibfield  {author} {\bibinfo {author} {\bibfnamefont {A.}~\bibnamefont {{Van
  Der Ziel}}},\ }\bibfield  {title} {\bibinfo {title} {{Unified presentation of
  1/f noise in electronic devices: Fundamental 1/f noise sources}},\ }\href
  {https://doi.org/10.1109/5.4401} {\bibfield  {journal} {\bibinfo  {journal}
  {Proc. IEEE}\ }\textbf {\bibinfo {volume} {76}},\ \bibinfo {pages} {233}
  (\bibinfo {year} {1988})}\BibitemShut {NoStop}%
\bibitem [{\citenamefont {Balandin}(2013)}]{Balandin2013}%
  \BibitemOpen
  \bibfield  {author} {\bibinfo {author} {\bibfnamefont {A.~A.}\ \bibnamefont
  {Balandin}},\ }\bibfield  {title} {\bibinfo {title} {{Low-frequency 1/f noise
  in graphene devices}},\ }\href {https://doi.org/10.1038/nnano.2013.144}
  {\bibfield  {journal} {\bibinfo  {journal} {Nat. Nanotechnol.}\ }\textbf
  {\bibinfo {volume} {8}},\ \bibinfo {pages} {549} (\bibinfo {year}
  {2013})}\BibitemShut {NoStop}%
\bibitem [{\citenamefont {Fleetwood}(2020)}]{Fleetwood2020}%
  \BibitemOpen
  \bibfield  {author} {\bibinfo {author} {\bibfnamefont {D.~M.}\ \bibnamefont
  {Fleetwood}},\ }\bibfield  {title} {\bibinfo {title} {{Origins of 1/f noise
  in electronic materials and devices: A historical perspective}},\ }in\ \href
  {https://doi.org/10.1007/978-3-030-37500-3_1} {\emph {\bibinfo {booktitle}
  {Noise Nanoscale Semicond. Devices}}},\ \bibinfo {editor} {edited by\
  \bibinfo {editor} {\bibfnamefont {T.}~\bibnamefont {Grasser}}}\ (\bibinfo
  {publisher} {Springer International Publishing},\ \bibinfo {address} {Cham},\
  \bibinfo {year} {2020})\ pp.\ \bibinfo {pages} {1--31}\BibitemShut {NoStop}%
\bibitem [{\citenamefont {Hossain}\ \emph
  {et~al.}(2011{\natexlab{a}})\citenamefont {Hossain}, \citenamefont
  {Rumyantsev}, \citenamefont {Teweldebrhan}, \citenamefont {Shahil},
  \citenamefont {Shur},\ and\ \citenamefont {Balandin}}]{Hossain2011}%
  \BibitemOpen
  \bibfield  {author} {\bibinfo {author} {\bibfnamefont {M.~Z.}\ \bibnamefont
  {Hossain}}, \bibinfo {author} {\bibfnamefont {S.~L.}\ \bibnamefont
  {Rumyantsev}}, \bibinfo {author} {\bibfnamefont {D.}~\bibnamefont
  {Teweldebrhan}}, \bibinfo {author} {\bibfnamefont {K.~M.}\ \bibnamefont
  {Shahil}}, \bibinfo {author} {\bibfnamefont {M.}~\bibnamefont {Shur}},\ and\
  \bibinfo {author} {\bibfnamefont {A.~A.}\ \bibnamefont {Balandin}},\
  }\bibfield  {title} {\bibinfo {title} {{1/F noise in conducting channels of
  topological insulator materials}},\ }\href
  {https://doi.org/10.1002/pssa.201026604} {\bibfield  {journal} {\bibinfo
  {journal} {Phys. Status Solidi Appl. Mater. Sci.}\ }\textbf {\bibinfo
  {volume} {208}},\ \bibinfo {pages} {144} (\bibinfo {year}
  {2011}{\natexlab{a}})}\BibitemShut {NoStop}%
\bibitem [{\citenamefont {Hossain}\ \emph
  {et~al.}(2011{\natexlab{b}})\citenamefont {Hossain}, \citenamefont
  {Rumyantsev}, \citenamefont {Shahil}, \citenamefont {Teweldebrhan},
  \citenamefont {Shur},\ and\ \citenamefont {Balandin}}]{Hossain2011a}%
  \BibitemOpen
  \bibfield  {author} {\bibinfo {author} {\bibfnamefont {M.~Z.}\ \bibnamefont
  {Hossain}}, \bibinfo {author} {\bibfnamefont {S.~L.}\ \bibnamefont
  {Rumyantsev}}, \bibinfo {author} {\bibfnamefont {K.~M.}\ \bibnamefont
  {Shahil}}, \bibinfo {author} {\bibfnamefont {D.}~\bibnamefont
  {Teweldebrhan}}, \bibinfo {author} {\bibfnamefont {M.}~\bibnamefont {Shur}},\
  and\ \bibinfo {author} {\bibfnamefont {A.~A.}\ \bibnamefont {Balandin}},\
  }\bibfield  {title} {\bibinfo {title} {{Low-frequency current fluctuations in
  "graphene-like" exfoliated thin-films of bismuth selenide topological
  insulators}},\ }\href {https://doi.org/10.1021/nn102861d} {\bibfield
  {journal} {\bibinfo  {journal} {ACS Nano}\ }\textbf {\bibinfo {volume} {5}},\
  \bibinfo {pages} {2657} (\bibinfo {year} {2011}{\natexlab{b}})}\BibitemShut
  {NoStop}%
\bibitem [{\citenamefont {Cascales}\ \emph {et~al.}(2015)\citenamefont
  {Cascales}, \citenamefont {Mart{\'{i}}nez}, \citenamefont {Katmis},
  \citenamefont {Chang}, \citenamefont {Guerrero}, \citenamefont {Moodera},\
  and\ \citenamefont {Aliev}}]{Cascales2015}%
  \BibitemOpen
  \bibfield  {author} {\bibinfo {author} {\bibfnamefont {J.~P.}\ \bibnamefont
  {Cascales}}, \bibinfo {author} {\bibfnamefont {I.}~\bibnamefont
  {Mart{\'{i}}nez}}, \bibinfo {author} {\bibfnamefont {F.}~\bibnamefont
  {Katmis}}, \bibinfo {author} {\bibfnamefont {C.~Z.}\ \bibnamefont {Chang}},
  \bibinfo {author} {\bibfnamefont {R.}~\bibnamefont {Guerrero}}, \bibinfo
  {author} {\bibfnamefont {J.~S.}\ \bibnamefont {Moodera}},\ and\ \bibinfo
  {author} {\bibfnamefont {F.~G.}\ \bibnamefont {Aliev}},\ }\bibfield  {title}
  {\bibinfo {title} {{Band structure of topological insulators from noise
  measurements in tunnel junctions}},\ }\href
  {https://doi.org/10.1063/1.4938243} {\bibfield  {journal} {\bibinfo
  {journal} {Appl. Phys. Lett.}\ }\textbf {\bibinfo {volume} {107}},\ \bibinfo
  {pages} {252402} (\bibinfo {year} {2015})}\BibitemShut {NoStop}%
\bibitem [{\citenamefont {Bhattacharyya}\ \emph {et~al.}(2015)\citenamefont
  {Bhattacharyya}, \citenamefont {Banerjee}, \citenamefont {Nhalil},
  \citenamefont {Islam}, \citenamefont {Dasgupta}, \citenamefont {Elizabeth},\
  and\ \citenamefont {Ghosh}}]{Bhattacharyya2015}%
  \BibitemOpen
  \bibfield  {author} {\bibinfo {author} {\bibfnamefont {S.}~\bibnamefont
  {Bhattacharyya}}, \bibinfo {author} {\bibfnamefont {M.}~\bibnamefont
  {Banerjee}}, \bibinfo {author} {\bibfnamefont {H.}~\bibnamefont {Nhalil}},
  \bibinfo {author} {\bibfnamefont {S.}~\bibnamefont {Islam}}, \bibinfo
  {author} {\bibfnamefont {C.}~\bibnamefont {Dasgupta}}, \bibinfo {author}
  {\bibfnamefont {S.}~\bibnamefont {Elizabeth}},\ and\ \bibinfo {author}
  {\bibfnamefont {A.}~\bibnamefont {Ghosh}},\ }\bibfield  {title} {\bibinfo
  {title} {{Bulk-induced 1/f noise at the surface of three-dimensional
  topological insulators}},\ }\href {https://doi.org/10.1021/acsnano.5b06163}
  {\bibfield  {journal} {\bibinfo  {journal} {ACS Nano}\ }\textbf {\bibinfo
  {volume} {9}},\ \bibinfo {pages} {12529} (\bibinfo {year}
  {2015})}\BibitemShut {NoStop}%
\bibitem [{\citenamefont {Bhattacharyya}\ \emph {et~al.}(2016)\citenamefont
  {Bhattacharyya}, \citenamefont {Kandala}, \citenamefont {Richardella},
  \citenamefont {Islam}, \citenamefont {Samarth},\ and\ \citenamefont
  {Ghosh}}]{Bhattacharyya2016}%
  \BibitemOpen
  \bibfield  {author} {\bibinfo {author} {\bibfnamefont {S.}~\bibnamefont
  {Bhattacharyya}}, \bibinfo {author} {\bibfnamefont {A.}~\bibnamefont
  {Kandala}}, \bibinfo {author} {\bibfnamefont {A.}~\bibnamefont
  {Richardella}}, \bibinfo {author} {\bibfnamefont {S.}~\bibnamefont {Islam}},
  \bibinfo {author} {\bibfnamefont {N.}~\bibnamefont {Samarth}},\ and\ \bibinfo
  {author} {\bibfnamefont {A.}~\bibnamefont {Ghosh}},\ }\bibfield  {title}
  {\bibinfo {title} {{Resistance noise in epitaxial thin films of ferromagnetic
  topological insulators}},\ }\href {https://doi.org/10.1063/1.4942412}
  {\bibfield  {journal} {\bibinfo  {journal} {Appl. Phys. Lett.}\ }\textbf
  {\bibinfo {volume} {108}},\ \bibinfo {pages} {82101} (\bibinfo {year}
  {2016})}\BibitemShut {NoStop}%
\bibitem [{\citenamefont {Zhang}\ \emph {et~al.}(2016)\citenamefont {Zhang},
  \citenamefont {Song}, \citenamefont {Feng}, \citenamefont {Ji},\ and\
  \citenamefont {Lu}}]{Zhang2016}%
  \BibitemOpen
  \bibfield  {author} {\bibinfo {author} {\bibfnamefont {H.}~\bibnamefont
  {Zhang}}, \bibinfo {author} {\bibfnamefont {Z.~J.}\ \bibnamefont {Song}},
  \bibinfo {author} {\bibfnamefont {J.~Y.}\ \bibnamefont {Feng}}, \bibinfo
  {author} {\bibfnamefont {Z.~Q.}\ \bibnamefont {Ji}},\ and\ \bibinfo {author}
  {\bibfnamefont {L.}~\bibnamefont {Lu}},\ }\bibfield  {title} {\bibinfo
  {title} {{Low-frequency noise in gate tunable topological insulator nanowire
  field emission transistor near the Dirac point}},\ }\href
  {https://doi.org/10.1088/0256-307X/33/8/087302} {\bibfield  {journal}
  {\bibinfo  {journal} {Chinese Phys. Lett.}\ }\textbf {\bibinfo {volume}
  {33}},\ \bibinfo {pages} {87302} (\bibinfo {year} {2016})}\BibitemShut
  {NoStop}%
\bibitem [{\citenamefont {Islam}\ \emph {et~al.}(2017)\citenamefont {Islam},
  \citenamefont {Bhattacharyya}, \citenamefont {Kandala}, \citenamefont
  {Richardella}, \citenamefont {Samarth},\ and\ \citenamefont
  {Ghosh}}]{Islam2017}%
  \BibitemOpen
  \bibfield  {author} {\bibinfo {author} {\bibfnamefont {S.}~\bibnamefont
  {Islam}}, \bibinfo {author} {\bibfnamefont {S.}~\bibnamefont
  {Bhattacharyya}}, \bibinfo {author} {\bibfnamefont {A.}~\bibnamefont
  {Kandala}}, \bibinfo {author} {\bibfnamefont {A.}~\bibnamefont
  {Richardella}}, \bibinfo {author} {\bibfnamefont {N.}~\bibnamefont
  {Samarth}},\ and\ \bibinfo {author} {\bibfnamefont {A.}~\bibnamefont
  {Ghosh}},\ }\bibfield  {title} {\bibinfo {title} {{Bulk-impurity induced
  noise in large-area epitaxial thin films of topological insulators}},\ }\href
  {https://doi.org/10.1063/1.4998464} {\bibfield  {journal} {\bibinfo
  {journal} {Appl. Phys. Lett.}\ }\textbf {\bibinfo {volume} {111}},\ \bibinfo
  {pages} {62107} (\bibinfo {year} {2017})}\BibitemShut {NoStop}%
\bibitem [{\citenamefont {Biswas}\ \emph {et~al.}(2017)\citenamefont {Biswas},
  \citenamefont {{Ciomaga Hatnean}}, \citenamefont {Balakrishnan},\ and\
  \citenamefont {Bid}}]{Biswas2017}%
  \BibitemOpen
  \bibfield  {author} {\bibinfo {author} {\bibfnamefont {S.}~\bibnamefont
  {Biswas}}, \bibinfo {author} {\bibfnamefont {M.}~\bibnamefont {{Ciomaga
  Hatnean}}}, \bibinfo {author} {\bibfnamefont {G.}~\bibnamefont
  {Balakrishnan}},\ and\ \bibinfo {author} {\bibfnamefont {A.}~\bibnamefont
  {Bid}},\ }\bibfield  {title} {\bibinfo {title} {{Probing the interplay
  between surface and bulk states in the topological Kondo insulator
  {SmB$_{6}$} through conductance fluctuation spectroscopy}},\ }\href
  {https://doi.org/10.1103/PhysRevB.95.205403} {\bibfield  {journal} {\bibinfo
  {journal} {Phys. Rev. B}\ }\textbf {\bibinfo {volume} {95}},\ \bibinfo
  {pages} {205403} (\bibinfo {year} {2017})}\BibitemShut {NoStop}%
\bibitem [{\citenamefont {Islam}\ \emph {et~al.}(2019)\citenamefont {Islam},
  \citenamefont {Bhattacharyya}, \citenamefont {Nhalil}, \citenamefont
  {Banerjee}, \citenamefont {Richardella}, \citenamefont {Kandala},
  \citenamefont {Sen}, \citenamefont {Samarth}, \citenamefont {Elizabeth},\
  and\ \citenamefont {Ghosh}}]{Islam2019}%
  \BibitemOpen
  \bibfield  {author} {\bibinfo {author} {\bibfnamefont {S.}~\bibnamefont
  {Islam}}, \bibinfo {author} {\bibfnamefont {S.}~\bibnamefont
  {Bhattacharyya}}, \bibinfo {author} {\bibfnamefont {H.}~\bibnamefont
  {Nhalil}}, \bibinfo {author} {\bibfnamefont {M.}~\bibnamefont {Banerjee}},
  \bibinfo {author} {\bibfnamefont {A.}~\bibnamefont {Richardella}}, \bibinfo
  {author} {\bibfnamefont {A.}~\bibnamefont {Kandala}}, \bibinfo {author}
  {\bibfnamefont {D.}~\bibnamefont {Sen}}, \bibinfo {author} {\bibfnamefont
  {N.}~\bibnamefont {Samarth}}, \bibinfo {author} {\bibfnamefont
  {S.}~\bibnamefont {Elizabeth}},\ and\ \bibinfo {author} {\bibfnamefont
  {A.}~\bibnamefont {Ghosh}},\ }\bibfield  {title} {\bibinfo {title}
  {{Low-temperature saturation of phase coherence length in topological
  insulators}},\ }\href {https://doi.org/10.1103/PhysRevB.99.245407} {\bibfield
   {journal} {\bibinfo  {journal} {Phys. Rev. B}\ }\textbf {\bibinfo {volume}
  {99}},\ \bibinfo {pages} {245407} (\bibinfo {year} {2019})}\BibitemShut
  {NoStop}%
\bibitem [{\citenamefont {Biswas}\ \emph {et~al.}(2019)\citenamefont {Biswas},
  \citenamefont {Gopal}, \citenamefont {Singh}, \citenamefont {Kant},
  \citenamefont {Mitra},\ and\ \citenamefont {Bid}}]{Biswas2019}%
  \BibitemOpen
  \bibfield  {author} {\bibinfo {author} {\bibfnamefont {S.}~\bibnamefont
  {Biswas}}, \bibinfo {author} {\bibfnamefont {R.~K.}\ \bibnamefont {Gopal}},
  \bibinfo {author} {\bibfnamefont {S.}~\bibnamefont {Singh}}, \bibinfo
  {author} {\bibfnamefont {R.}~\bibnamefont {Kant}}, \bibinfo {author}
  {\bibfnamefont {C.}~\bibnamefont {Mitra}},\ and\ \bibinfo {author}
  {\bibfnamefont {A.}~\bibnamefont {Bid}},\ }\bibfield  {title} {\bibinfo
  {title} {{Resistance fluctuation spectroscopy of thin films of 3D topological
  insulator {BiSbTeSe$_{1.6}$}}},\ }\href {https://doi.org/10.1063/1.5119288}
  {\bibfield  {journal} {\bibinfo  {journal} {Appl. Phys. Lett.}\ }\textbf
  {\bibinfo {volume} {115}},\ \bibinfo {pages} {131601} (\bibinfo {year}
  {2019})}\BibitemShut {NoStop}%
\bibitem [{\citenamefont {Kunakova}\ \emph {et~al.}(2019)\citenamefont
  {Kunakova}, \citenamefont {Bauch}, \citenamefont {Trabaldo}, \citenamefont
  {Andzane}, \citenamefont {Erts},\ and\ \citenamefont
  {Lombardi}}]{Kunakova2019}%
  \BibitemOpen
  \bibfield  {author} {\bibinfo {author} {\bibfnamefont {G.}~\bibnamefont
  {Kunakova}}, \bibinfo {author} {\bibfnamefont {T.}~\bibnamefont {Bauch}},
  \bibinfo {author} {\bibfnamefont {E.}~\bibnamefont {Trabaldo}}, \bibinfo
  {author} {\bibfnamefont {J.}~\bibnamefont {Andzane}}, \bibinfo {author}
  {\bibfnamefont {D.}~\bibnamefont {Erts}},\ and\ \bibinfo {author}
  {\bibfnamefont {F.}~\bibnamefont {Lombardi}},\ }\bibfield  {title} {\bibinfo
  {title} {{High transparency {Bi$_{2}$Se$_{3}$} topological insulator
  nanoribbon Josephson junctions with low resistive noise properties}},\ }\href
  {https://doi.org/10.1063/1.5123554} {\bibfield  {journal} {\bibinfo
  {journal} {Appl. Phys. Lett.}\ }\textbf {\bibinfo {volume} {115}},\ \bibinfo
  {pages} {172601} (\bibinfo {year} {2019})}\BibitemShut {NoStop}%
\bibitem [{\citenamefont {Islam}\ \emph {et~al.}(2020)\citenamefont {Islam},
  \citenamefont {Bhattacharyya}, \citenamefont {Nhalil}, \citenamefont
  {Elizabeth},\ and\ \citenamefont {Ghosh}}]{Islam2020}%
  \BibitemOpen
  \bibfield  {author} {\bibinfo {author} {\bibfnamefont {S.}~\bibnamefont
  {Islam}}, \bibinfo {author} {\bibfnamefont {S.}~\bibnamefont
  {Bhattacharyya}}, \bibinfo {author} {\bibfnamefont {H.}~\bibnamefont
  {Nhalil}}, \bibinfo {author} {\bibfnamefont {S.}~\bibnamefont {Elizabeth}},\
  and\ \bibinfo {author} {\bibfnamefont {A.}~\bibnamefont {Ghosh}},\ }\bibfield
   {title} {\bibinfo {title} {{Signature of pseudodiffusive transport in
  mesoscopic topological insulators}},\ }\href
  {https://doi.org/10.1103/PhysRevResearch.2.033019} {\bibfield  {journal}
  {\bibinfo  {journal} {Phys. Rev. Res.}\ }\textbf {\bibinfo {volume} {2}},\
  \bibinfo {pages} {33019} (\bibinfo {year} {2020})}\BibitemShut {NoStop}%
\bibitem [{\citenamefont {Kane}\ and\ \citenamefont {Mele}(2005)}]{Kane2005}%
  \BibitemOpen
  \bibfield  {author} {\bibinfo {author} {\bibfnamefont {C.~L.}\ \bibnamefont
  {Kane}}\ and\ \bibinfo {author} {\bibfnamefont {E.~J.}\ \bibnamefont
  {Mele}},\ }\bibfield  {title} {\bibinfo {title} {{{Z$_{2}$} topological order
  and the quantum spin Hall effect}},\ }\href
  {https://doi.org/10.1103/PhysRevLett.95.146802} {\bibfield  {journal}
  {\bibinfo  {journal} {Phys. Rev. Lett.}\ }\textbf {\bibinfo {volume} {95}},\
  \bibinfo {pages} {146802} (\bibinfo {year} {2005})}\BibitemShut {NoStop}%
\bibitem [{\citenamefont {Bernevig}\ \emph {et~al.}(2006)\citenamefont
  {Bernevig}, \citenamefont {Hughes},\ and\ \citenamefont
  {Zhang}}]{Bernevig2006}%
  \BibitemOpen
  \bibfield  {author} {\bibinfo {author} {\bibfnamefont {B.~A.}\ \bibnamefont
  {Bernevig}}, \bibinfo {author} {\bibfnamefont {T.~L.}\ \bibnamefont
  {Hughes}},\ and\ \bibinfo {author} {\bibfnamefont {S.~C.}\ \bibnamefont
  {Zhang}},\ }\bibfield  {title} {\bibinfo {title} {{Quantum spin Hall effect
  and topological phase transition in HgTe quantum wells}},\ }\href
  {https://doi.org/10.1126/science.1133734} {\bibfield  {journal} {\bibinfo
  {journal} {Science}\ }\textbf {\bibinfo {volume} {314}},\ \bibinfo {pages}
  {1757} (\bibinfo {year} {2006})}\BibitemShut {NoStop}%
\bibitem [{\citenamefont {K{\"{o}}nig}\ \emph {et~al.}(2007)\citenamefont
  {K{\"{o}}nig}, \citenamefont {Wiedmann}, \citenamefont {Br{\"{u}}ne},
  \citenamefont {Roth}, \citenamefont {Buhmann}, \citenamefont {Molenkamp},
  \citenamefont {Qi},\ and\ \citenamefont {Zhang}}]{Konig2007}%
  \BibitemOpen
  \bibfield  {author} {\bibinfo {author} {\bibfnamefont {M.}~\bibnamefont
  {K{\"{o}}nig}}, \bibinfo {author} {\bibfnamefont {S.}~\bibnamefont
  {Wiedmann}}, \bibinfo {author} {\bibfnamefont {C.}~\bibnamefont
  {Br{\"{u}}ne}}, \bibinfo {author} {\bibfnamefont {A.}~\bibnamefont {Roth}},
  \bibinfo {author} {\bibfnamefont {H.}~\bibnamefont {Buhmann}}, \bibinfo
  {author} {\bibfnamefont {L.~W.}\ \bibnamefont {Molenkamp}}, \bibinfo {author}
  {\bibfnamefont {X.~L.}\ \bibnamefont {Qi}},\ and\ \bibinfo {author}
  {\bibfnamefont {S.~C.}\ \bibnamefont {Zhang}},\ }\bibfield  {title} {\bibinfo
  {title} {{Quantum spin Hall insulator state in HgTe quantum wells}},\ }\href
  {https://doi.org/10.1126/science.1148047} {\bibfield  {journal} {\bibinfo
  {journal} {Science}\ }\textbf {\bibinfo {volume} {318}},\ \bibinfo {pages}
  {766} (\bibinfo {year} {2007})}\BibitemShut {NoStop}%
\bibitem [{\citenamefont {Fu}\ \emph {et~al.}(2007)\citenamefont {Fu},
  \citenamefont {Kane},\ and\ \citenamefont {Mele}}]{Fu2007}%
  \BibitemOpen
  \bibfield  {author} {\bibinfo {author} {\bibfnamefont {L.}~\bibnamefont
  {Fu}}, \bibinfo {author} {\bibfnamefont {C.~L.}\ \bibnamefont {Kane}},\ and\
  \bibinfo {author} {\bibfnamefont {E.~J.}\ \bibnamefont {Mele}},\ }\bibfield
  {title} {\bibinfo {title} {{Topological insulators in three dimensions}},\
  }\href {https://doi.org/10.1103/PhysRevLett.98.106803} {\bibfield  {journal}
  {\bibinfo  {journal} {Phys. Rev. Lett.}\ }\textbf {\bibinfo {volume} {98}},\
  \bibinfo {pages} {106803} (\bibinfo {year} {2007})}\BibitemShut {NoStop}%
\bibitem [{\citenamefont {Zhang}\ \emph {et~al.}(2009)\citenamefont {Zhang},
  \citenamefont {Liu}, \citenamefont {Qi}, \citenamefont {Dai}, \citenamefont
  {Fang},\ and\ \citenamefont {Zhang}}]{Zhang2009}%
  \BibitemOpen
  \bibfield  {author} {\bibinfo {author} {\bibfnamefont {H.}~\bibnamefont
  {Zhang}}, \bibinfo {author} {\bibfnamefont {C.~X.}\ \bibnamefont {Liu}},
  \bibinfo {author} {\bibfnamefont {X.~L.}\ \bibnamefont {Qi}}, \bibinfo
  {author} {\bibfnamefont {X.}~\bibnamefont {Dai}}, \bibinfo {author}
  {\bibfnamefont {Z.}~\bibnamefont {Fang}},\ and\ \bibinfo {author}
  {\bibfnamefont {S.~C.}\ \bibnamefont {Zhang}},\ }\bibfield  {title} {\bibinfo
  {title} {{Topological insulators in {Bi$_2$Se$_3$}, {Bi$_2$Te$_3$} and
  {Sb$_2$Te$_3$} with a single Dirac cone on the surface}},\ }\href
  {https://doi.org/10.1038/nphys1270} {\bibfield  {journal} {\bibinfo
  {journal} {Nat. Phys.}\ }\textbf {\bibinfo {volume} {5}},\ \bibinfo {pages}
  {438} (\bibinfo {year} {2009})}\BibitemShut {NoStop}%
\bibitem [{\citenamefont {Moore}(2010)}]{Moore2010}%
  \BibitemOpen
  \bibfield  {author} {\bibinfo {author} {\bibfnamefont {J.~E.}\ \bibnamefont
  {Moore}},\ }\bibfield  {title} {\bibinfo {title} {{The birth of topological
  insulators}},\ }\href {https://doi.org/10.1038/nature08916} {\bibfield
  {journal} {\bibinfo  {journal} {Nature}\ }\textbf {\bibinfo {volume} {464}},\
  \bibinfo {pages} {194} (\bibinfo {year} {2010})}\BibitemShut {NoStop}%
\bibitem [{\citenamefont {Hasan}\ and\ \citenamefont {Kane}(2010)}]{Hasan2010}%
  \BibitemOpen
  \bibfield  {author} {\bibinfo {author} {\bibfnamefont {M.~Z.}\ \bibnamefont
  {Hasan}}\ and\ \bibinfo {author} {\bibfnamefont {C.~L.}\ \bibnamefont
  {Kane}},\ }\bibfield  {title} {\bibinfo {title} {{Colloquium: Topological
  insulators}},\ }\href {https://doi.org/10.1103/RevModPhys.82.3045} {\bibfield
   {journal} {\bibinfo  {journal} {Rev. Mod. Phys.}\ }\textbf {\bibinfo
  {volume} {82}},\ \bibinfo {pages} {3045} (\bibinfo {year}
  {2010})}\BibitemShut {NoStop}%
\bibitem [{\citenamefont {Ando}(2013)}]{Ando2013}%
  \BibitemOpen
  \bibfield  {author} {\bibinfo {author} {\bibfnamefont {Y.}~\bibnamefont
  {Ando}},\ }\bibfield  {title} {\bibinfo {title} {{Topological insulator
  materials}},\ }\href {https://doi.org/10.7566/JPSJ.82.102001} {\bibfield
  {journal} {\bibinfo  {journal} {J. Phys. Soc. Japan}\ }\textbf {\bibinfo
  {volume} {82}},\ \bibinfo {pages} {102001} (\bibinfo {year}
  {2013})}\BibitemShut {NoStop}%
\bibitem [{\citenamefont {Paladino}\ \emph {et~al.}(2014)\citenamefont
  {Paladino}, \citenamefont {Galperin}, \citenamefont {Falci},\ and\
  \citenamefont {Altshuler}}]{Paladino2014}%
  \BibitemOpen
  \bibfield  {author} {\bibinfo {author} {\bibfnamefont {E.}~\bibnamefont
  {Paladino}}, \bibinfo {author} {\bibfnamefont {Y.}~\bibnamefont {Galperin}},
  \bibinfo {author} {\bibfnamefont {G.}~\bibnamefont {Falci}},\ and\ \bibinfo
  {author} {\bibfnamefont {B.~L.}\ \bibnamefont {Altshuler}},\ }\bibfield
  {title} {\bibinfo {title} {{1/ f noise: Implications for solid-state quantum
  information}},\ }\href {https://doi.org/10.1103/RevModPhys.86.361} {\bibfield
   {journal} {\bibinfo  {journal} {Rev. Mod. Phys.}\ }\textbf {\bibinfo
  {volume} {86}},\ \bibinfo {pages} {361} (\bibinfo {year} {2014})}\BibitemShut
  {NoStop}%
\bibitem [{\citenamefont {Arute}\ \emph {et~al.}(2019)\citenamefont {Arute}
  \emph {et~al.}}]{Arute2019}%
  \BibitemOpen
  \bibfield  {author} {\bibinfo {author} {\bibfnamefont {F.}~\bibnamefont
  {Arute}} \emph {et~al.},\ }\bibfield  {title} {\bibinfo {title} {{Quantum
  supremacy using a programmable superconducting processor}},\ }\href
  {https://doi.org/10.1038/s41586-019-1666-5} {\bibfield  {journal} {\bibinfo
  {journal} {Nature}\ }\textbf {\bibinfo {volume} {574}},\ \bibinfo {pages}
  {505} (\bibinfo {year} {2019})}\BibitemShut {NoStop}%
\bibitem [{\citenamefont {Carruthers}(1971)}]{Carruthers1971}%
  \BibitemOpen
  \bibfield  {author} {\bibinfo {author} {\bibfnamefont {T.}~\bibnamefont
  {Carruthers}},\ }\bibfield  {title} {\bibinfo {title} {{Bias-dependent
  structure in excess noise in GaAs schottky tunnel junctions}},\ }\href
  {https://doi.org/10.1063/1.1653469} {\bibfield  {journal} {\bibinfo
  {journal} {Appl. Phys. Lett.}\ }\textbf {\bibinfo {volume} {18}},\ \bibinfo
  {pages} {35} (\bibinfo {year} {1971})}\BibitemShut {NoStop}%
\bibitem [{\citenamefont {Mihaila}(1984)}]{Mihaila1984}%
  \BibitemOpen
  \bibfield  {author} {\bibinfo {author} {\bibfnamefont {M.}~\bibnamefont
  {Mihaila}},\ }\bibfield  {title} {\bibinfo {title} {{Phonon observations from
  1/f noise measurements}},\ }\href
  {https://doi.org/10.1016/0375-9601(84)90366-9} {\bibfield  {journal}
  {\bibinfo  {journal} {Phys. Lett. A}\ }\textbf {\bibinfo {volume} {104}},\
  \bibinfo {pages} {157} (\bibinfo {year} {1984})}\BibitemShut {NoStop}%
\bibitem [{\citenamefont {Mihaila}(2004)}]{Mihaila2004}%
  \BibitemOpen
  \bibfield  {author} {\bibinfo {author} {\bibfnamefont {M.~N.}\ \bibnamefont
  {Mihaila}},\ }\bibfield  {title} {\bibinfo {title} {{Phonon-induced 1/f noise
  in MOS transistors}},\ }\href {https://doi.org/10.1142/S0219477504001938}
  {\bibfield  {journal} {\bibinfo  {journal} {Fluct. Noise Lett.}\ }\textbf
  {\bibinfo {volume} {4}},\ \bibinfo {pages} {L329} (\bibinfo {year}
  {2004})}\BibitemShut {NoStop}%
\bibitem [{\citenamefont {Yanson}\ \emph {et~al.}(1982)\citenamefont {Yanson},
  \citenamefont {Akimenko},\ and\ \citenamefont {Verkin}}]{Yanson1982}%
  \BibitemOpen
  \bibfield  {author} {\bibinfo {author} {\bibfnamefont {I.~K.}\ \bibnamefont
  {Yanson}}, \bibinfo {author} {\bibfnamefont {A.~I.}\ \bibnamefont
  {Akimenko}},\ and\ \bibinfo {author} {\bibfnamefont {A.~B.}\ \bibnamefont
  {Verkin}},\ }\bibfield  {title} {\bibinfo {title} {{Electrical fluctuations
  in normal metal point-contacts}},\ }\href
  {https://doi.org/10.1016/0038-1098(82)90988-7} {\bibfield  {journal}
  {\bibinfo  {journal} {Solid State Commun.}\ }\textbf {\bibinfo {volume}
  {43}},\ \bibinfo {pages} {765} (\bibinfo {year} {1982})}\BibitemShut
  {NoStop}%
\bibitem [{\citenamefont {Akimenko}\ \emph {et~al.}(1984)\citenamefont
  {Akimenko}, \citenamefont {Verkin},\ and\ \citenamefont
  {Yanson}}]{Akimenko1984}%
  \BibitemOpen
  \bibfield  {author} {\bibinfo {author} {\bibfnamefont {A.~I.}\ \bibnamefont
  {Akimenko}}, \bibinfo {author} {\bibfnamefont {A.~B.}\ \bibnamefont
  {Verkin}},\ and\ \bibinfo {author} {\bibfnamefont {I.~K.}\ \bibnamefont
  {Yanson}},\ }\bibfield  {title} {\bibinfo {title} {{Point-contact noise
  spectroscopy of phonons in metals}},\ }\href
  {https://doi.org/10.1007/BF00683277} {\bibfield  {journal} {\bibinfo
  {journal} {J. Low Temp. Phys.}\ }\textbf {\bibinfo {volume} {54}},\ \bibinfo
  {pages} {247} (\bibinfo {year} {1984})}\BibitemShut {NoStop}%
\bibitem [{\citenamefont {Mihaila}(1985)}]{Mihaila1985}%
  \BibitemOpen
  \bibfield  {author} {\bibinfo {author} {\bibfnamefont {M.}~\bibnamefont
  {Mihaila}},\ }\bibfield  {title} {\bibinfo {title} {{Phonon signatures in the
  1/f noise parameter of copper, silver and silicon}},\ }\href
  {https://doi.org/10.1016/0375-9601(85)90426-8} {\bibfield  {journal}
  {\bibinfo  {journal} {Phys. Lett. A}\ }\textbf {\bibinfo {volume} {107}},\
  \bibinfo {pages} {465} (\bibinfo {year} {1985})}\BibitemShut {NoStop}%
\bibitem [{\citenamefont {Mihaila}(2007)}]{Mihaila2000}%
  \BibitemOpen
  \bibfield  {author} {\bibinfo {author} {\bibfnamefont {M.~N.}\ \bibnamefont
  {Mihaila}},\ }\bibfield  {title} {\bibinfo {title} {{Phonon Fine Structure in
  the 1/f Noise of Metals, Semiconductors and Semiconductor Devices}},\ }in\
  \href {https://doi.org/10.1007/3-540-45463-2_11} {\emph {\bibinfo {booktitle}
  {Noise, Oscil. Algebr. Randomness}}},\ \bibinfo {series and number} {Lecture
  Notes in Physics},\ \bibinfo {editor} {edited by\ \bibinfo {editor}
  {\bibfnamefont {M.}~\bibnamefont {Planat}}}\ (\bibinfo  {publisher}
  {Springer},\ \bibinfo {address} {Berlin, Heidelberg},\ \bibinfo {year}
  {2007})\ pp.\ \bibinfo {pages} {216--231}\BibitemShut {NoStop}%
\bibitem [{\citenamefont {Planat}\ and\ \citenamefont
  {Gagnepain}(1987)}]{Planat1987}%
  \BibitemOpen
  \bibfield  {author} {\bibinfo {author} {\bibfnamefont {M.}~\bibnamefont
  {Planat}}\ and\ \bibinfo {author} {\bibfnamefont {J.~J.}\ \bibnamefont
  {Gagnepain}},\ }\bibfield  {title} {\bibinfo {title} {{1/f noise in quartz
  crystal resonators in relation to acoustic losses and frequency
  fispersion}},\ }\href {https://doi.org/10.1063/1.98143} {\bibfield  {journal}
  {\bibinfo  {journal} {Appl. Phys. Lett.}\ }\textbf {\bibinfo {volume} {50}},\
  \bibinfo {pages} {510} (\bibinfo {year} {1987})}\BibitemShut {NoStop}%
\bibitem [{\citenamefont {Back}\ \emph {et~al.}(2009)\citenamefont {Back},
  \citenamefont {Tsai}, \citenamefont {Kim}, \citenamefont {Mohammadi},\ and\
  \citenamefont {Shim}}]{Back2009}%
  \BibitemOpen
  \bibfield  {author} {\bibinfo {author} {\bibfnamefont {J.~H.}\ \bibnamefont
  {Back}}, \bibinfo {author} {\bibfnamefont {C.~L.}\ \bibnamefont {Tsai}},
  \bibinfo {author} {\bibfnamefont {S.}~\bibnamefont {Kim}}, \bibinfo {author}
  {\bibfnamefont {S.}~\bibnamefont {Mohammadi}},\ and\ \bibinfo {author}
  {\bibfnamefont {M.}~\bibnamefont {Shim}},\ }\bibfield  {title} {\bibinfo
  {title} {{Manifestation of Kohn anomaly in 1/f fluctuations in metallic
  carbon nanotubes}},\ }\href {https://doi.org/10.1103/PhysRevLett.103.215501}
  {\bibfield  {journal} {\bibinfo  {journal} {Phys. Rev. Lett.}\ }\textbf
  {\bibinfo {volume} {103}},\ \bibinfo {pages} {215501} (\bibinfo {year}
  {2009})}\BibitemShut {NoStop}%
\bibitem [{\citenamefont {Mihaila}(2009)}]{Mihaila_patent}%
  \BibitemOpen
  \bibfield  {author} {\bibinfo {author} {\bibfnamefont {M.}~\bibnamefont
  {Mihaila}},\ }\href@noop {} {\bibinfo {title} {System of phonon spectroscopy,
  {U.S. Patent 7612551B2, Nov. 3 }}} (\bibinfo {year} {2009})\BibitemShut
  {NoStop}%
\bibitem [{\citenamefont {Mihaila}\ \emph {et~al.}(2019)\citenamefont
  {Mihaila}, \citenamefont {Ursutiu},\ and\ \citenamefont
  {Sandu}}]{Mihaila2019}%
  \BibitemOpen
  \bibfield  {author} {\bibinfo {author} {\bibfnamefont {M.}~\bibnamefont
  {Mihaila}}, \bibinfo {author} {\bibfnamefont {D.}~\bibnamefont {Ursutiu}},\
  and\ \bibinfo {author} {\bibfnamefont {I.}~\bibnamefont {Sandu}},\ }\bibfield
   {title} {\bibinfo {title} {{Electron-phonon coupling as the source of 1/f
  noise in carbon soot}},\ }\href {https://doi.org/10.1038/s41598-018-36544-4}
  {\bibfield  {journal} {\bibinfo  {journal} {Sci. Rep.}\ }\textbf {\bibinfo
  {volume} {9}},\ \bibinfo {pages} {947} (\bibinfo {year} {2019})}\BibitemShut
  {NoStop}%
\bibitem [{\citenamefont {Mihaila}(2011)}]{Mihaila2011}%
  \BibitemOpen
  \bibfield  {author} {\bibinfo {author} {\bibfnamefont {M.~N.}\ \bibnamefont
  {Mihaila}},\ }\bibfield  {title} {\bibinfo {title} {{Atomic vibration-induced
  1/f noise in sensing nanomaterials}},\ }in\ \href
  {https://doi.org/10.1109/ICNF.2011.5994385} {\emph {\bibinfo {booktitle}
  {Proc. IEEE 21st Int. Conf. Noise Fluctuations, ICNF 2011}}},\ \bibinfo
  {editor} {edited by\ \bibinfo {editor} {\bibfnamefont {J.}~\bibnamefont
  {Deen}}}\ (\bibinfo {year} {2011})\ pp.\ \bibinfo {pages}
  {61--64}\BibitemShut {NoStop}%
\bibitem [{\citenamefont {Tsutsui}\ \emph {et~al.}(2010)\citenamefont
  {Tsutsui}, \citenamefont {Taniguchi},\ and\ \citenamefont
  {Kawai}}]{Tsutsui2010}%
  \BibitemOpen
  \bibfield  {author} {\bibinfo {author} {\bibfnamefont {M.}~\bibnamefont
  {Tsutsui}}, \bibinfo {author} {\bibfnamefont {M.}~\bibnamefont {Taniguchi}},\
  and\ \bibinfo {author} {\bibfnamefont {T.}~\bibnamefont {Kawai}},\ }\bibfield
   {title} {\bibinfo {title} {{Single-molecule identification via electric
  current noise}},\ }\href {https://doi.org/10.1038/ncomms1141} {\bibfield
  {journal} {\bibinfo  {journal} {Nat. Commun.}\ }\textbf {\bibinfo {volume}
  {1}},\ \bibinfo {pages} {138} (\bibinfo {year} {2010})}\BibitemShut {NoStop}%
\bibitem [{\citenamefont {Kim}\ and\ \citenamefont {Song}(2021)}]{Kim2021}%
  \BibitemOpen
  \bibfield  {author} {\bibinfo {author} {\bibfnamefont {Y.}~\bibnamefont
  {Kim}}\ and\ \bibinfo {author} {\bibfnamefont {H.}~\bibnamefont {Song}},\
  }\bibfield  {title} {\bibinfo {title} {{Noise spectroscopy of molecular
  electronic junctions}},\ }\href {https://doi.org/10.1063/5.0027602}
  {\bibfield  {journal} {\bibinfo  {journal} {Appl. Phys. Rev.}\ }\textbf
  {\bibinfo {volume} {8}},\ \bibinfo {pages} {11303} (\bibinfo {year}
  {2021})}\BibitemShut {NoStop}%
\bibitem [{\citenamefont {Mihaila}(2014)}]{Mihaila1999}%
  \BibitemOpen
  \bibfield  {author} {\bibinfo {author} {\bibfnamefont {M.~N.}\ \bibnamefont
  {Mihaila}},\ }\bibfield  {title} {\bibinfo {title} {{Possible connection
  between 1/f noise parameter and the Eliashberg function}},\ }in\ \href
  {https://doi.org/10.1063/1.58284} {\emph {\bibinfo {booktitle} {AIP Conf.
  Proc.}}}\ (\bibinfo {year} {2014})\ pp.\ \bibinfo {pages}
  {48--55}\BibitemShut {NoStop}%
\bibitem [{\citenamefont {Klobes}\ \emph {et~al.}(2015)\citenamefont {Klobes},
  \citenamefont {Bessas}, \citenamefont {Juranyi}, \citenamefont
  {G{\"{o}}rlitz}, \citenamefont {Pacheco},\ and\ \citenamefont
  {Hermann}}]{Klobes2015}%
  \BibitemOpen
  \bibfield  {author} {\bibinfo {author} {\bibfnamefont {B.}~\bibnamefont
  {Klobes}}, \bibinfo {author} {\bibfnamefont {D.}~\bibnamefont {Bessas}},
  \bibinfo {author} {\bibfnamefont {F.}~\bibnamefont {Juranyi}}, \bibinfo
  {author} {\bibfnamefont {H.}~\bibnamefont {G{\"{o}}rlitz}}, \bibinfo {author}
  {\bibfnamefont {V.}~\bibnamefont {Pacheco}},\ and\ \bibinfo {author}
  {\bibfnamefont {R.~P.}\ \bibnamefont {Hermann}},\ }\bibfield  {title}
  {\bibinfo {title} {{Effect of nanocrystallinity on lattice dynamics in
  {Bi$_2$Te$_3$} based thermoelectrics}},\ }\href
  {https://doi.org/10.1002/pssr.201409479} {\bibfield  {journal} {\bibinfo
  {journal} {Phys. Status Solidi - Rapid Res. Lett.}\ }\textbf {\bibinfo
  {volume} {9}},\ \bibinfo {pages} {57} (\bibinfo {year} {2015})}\BibitemShut
  {NoStop}%
\bibitem [{\citenamefont {Xie}\ \emph {et~al.}(2010)\citenamefont {Xie},
  \citenamefont {He}, \citenamefont {Kang}, \citenamefont {Tang}, \citenamefont
  {Zhu}, \citenamefont {Laver}, \citenamefont {Wang}, \citenamefont {Copley},
  \citenamefont {Brown}, \citenamefont {Zhang},\ and\ \citenamefont
  {Tritt}}]{Xie2010}%
  \BibitemOpen
  \bibfield  {author} {\bibinfo {author} {\bibfnamefont {W.}~\bibnamefont
  {Xie}}, \bibinfo {author} {\bibfnamefont {J.}~\bibnamefont {He}}, \bibinfo
  {author} {\bibfnamefont {H.~J.}\ \bibnamefont {Kang}}, \bibinfo {author}
  {\bibfnamefont {X.}~\bibnamefont {Tang}}, \bibinfo {author} {\bibfnamefont
  {S.}~\bibnamefont {Zhu}}, \bibinfo {author} {\bibfnamefont {M.}~\bibnamefont
  {Laver}}, \bibinfo {author} {\bibfnamefont {S.}~\bibnamefont {Wang}},
  \bibinfo {author} {\bibfnamefont {J.~R.}\ \bibnamefont {Copley}}, \bibinfo
  {author} {\bibfnamefont {C.~M.}\ \bibnamefont {Brown}}, \bibinfo {author}
  {\bibfnamefont {Q.}~\bibnamefont {Zhang}},\ and\ \bibinfo {author}
  {\bibfnamefont {T.~M.}\ \bibnamefont {Tritt}},\ }\bibfield  {title} {\bibinfo
  {title} {{Identifying the specific nanostructures responsible for the high
  thermoelectric performance of {(Bi,Sb)$_2$Te$_3$} nanocomposites}},\ }\href
  {https://doi.org/10.1021/nl100804a} {\bibfield  {journal} {\bibinfo
  {journal} {Nano Lett.}\ }\textbf {\bibinfo {volume} {10}},\ \bibinfo {pages}
  {3283} (\bibinfo {year} {2010})}\BibitemShut {NoStop}%
\bibitem [{\citenamefont {German}\ \emph {et~al.}(2019)\citenamefont {German},
  \citenamefont {Komleva}, \citenamefont {Stein}, \citenamefont {Mazurenko},
  \citenamefont {Wang}, \citenamefont {Streltsov}, \citenamefont {Ando},\ and\
  \citenamefont {{Van Loosdrecht}}}]{German2019}%
  \BibitemOpen
  \bibfield  {author} {\bibinfo {author} {\bibfnamefont {R.}~\bibnamefont
  {German}}, \bibinfo {author} {\bibfnamefont {E.~V.}\ \bibnamefont {Komleva}},
  \bibinfo {author} {\bibfnamefont {P.}~\bibnamefont {Stein}}, \bibinfo
  {author} {\bibfnamefont {V.~G.}\ \bibnamefont {Mazurenko}}, \bibinfo {author}
  {\bibfnamefont {Z.}~\bibnamefont {Wang}}, \bibinfo {author} {\bibfnamefont
  {S.~V.}\ \bibnamefont {Streltsov}}, \bibinfo {author} {\bibfnamefont
  {Y.}~\bibnamefont {Ando}},\ and\ \bibinfo {author} {\bibfnamefont {P.~H.}\
  \bibnamefont {{Van Loosdrecht}}},\ }\bibfield  {title} {\bibinfo {title}
  {{Phonon mode calculations and Raman spectroscopy of the bulk-insulating
  topological insulator {BiSbTeSe$_2$}}},\ }\href
  {https://doi.org/10.1103/PhysRevMaterials.3.054204} {\bibfield  {journal}
  {\bibinfo  {journal} {Phys. Rev. Mater.}\ }\textbf {\bibinfo {volume} {3}},\
  \bibinfo {pages} {54204} (\bibinfo {year} {2019})}\BibitemShut {NoStop}%
\bibitem [{\citenamefont {Hooge}(1969)}]{Hooge1969}%
  \BibitemOpen
  \bibfield  {author} {\bibinfo {author} {\bibfnamefont {F.~N.}\ \bibnamefont
  {Hooge}},\ }\bibfield  {title} {\bibinfo {title} {{1/f noise is no surface
  effect}},\ }\href {https://doi.org/10.1016/0375-9601(69)90076-0} {\bibfield
  {journal} {\bibinfo  {journal} {Phys. Lett. A}\ }\textbf {\bibinfo {volume}
  {29}},\ \bibinfo {pages} {139} (\bibinfo {year} {1969})}\BibitemShut
  {NoStop}%
\bibitem [{\citenamefont {Wagner}\ \emph {et~al.}(1978)\citenamefont {Wagner},
  \citenamefont {Dolling}, \citenamefont {Powell},\ and\ \citenamefont
  {Landwehr}}]{Wagner1978}%
  \BibitemOpen
  \bibfield  {author} {\bibinfo {author} {\bibfnamefont {V.}~\bibnamefont
  {Wagner}}, \bibinfo {author} {\bibfnamefont {G.}~\bibnamefont {Dolling}},
  \bibinfo {author} {\bibfnamefont {B.~M.}\ \bibnamefont {Powell}},\ and\
  \bibinfo {author} {\bibfnamefont {G.}~\bibnamefont {Landwehr}},\ }\bibfield
  {title} {\bibinfo {title} {{Lattice vibrations of {Bi$_2$Te$_3$}}},\ }\href
  {https://doi.org/10.1002/pssb.2220850134} {\bibfield  {journal} {\bibinfo
  {journal} {Phys. Status Solidi}\ }\textbf {\bibinfo {volume} {85}},\ \bibinfo
  {pages} {311} (\bibinfo {year} {1978})}\BibitemShut {NoStop}%
\bibitem [{\citenamefont {Kullmann}\ \emph {et~al.}(1990)\citenamefont
  {Kullmann}, \citenamefont {Eichhorn}, \citenamefont {Rauh}, \citenamefont
  {Geick}, \citenamefont {Eckold},\ and\ \citenamefont
  {Steigenberger}}]{Kullmann1990}%
  \BibitemOpen
  \bibfield  {author} {\bibinfo {author} {\bibfnamefont {W.}~\bibnamefont
  {Kullmann}}, \bibinfo {author} {\bibfnamefont {G.}~\bibnamefont {Eichhorn}},
  \bibinfo {author} {\bibfnamefont {H.}~\bibnamefont {Rauh}}, \bibinfo {author}
  {\bibfnamefont {R.}~\bibnamefont {Geick}}, \bibinfo {author} {\bibfnamefont
  {G.}~\bibnamefont {Eckold}},\ and\ \bibinfo {author} {\bibfnamefont
  {U.}~\bibnamefont {Steigenberger}},\ }\bibfield  {title} {\bibinfo {title}
  {{Lattice dynamics and phonon dispersion in the narrow gap semiconductor
  {Bi$_2$Te$_3$} with sandwich structure}},\ }\href
  {https://doi.org/10.1002/pssb.2221620109} {\bibfield  {journal} {\bibinfo
  {journal} {Phys. Status Solidi}\ }\textbf {\bibinfo {volume} {162}},\
  \bibinfo {pages} {125} (\bibinfo {year} {1990})}\BibitemShut {NoStop}%
\bibitem [{\citenamefont {Shahil}\ \emph {et~al.}(2010)\citenamefont {Shahil},
  \citenamefont {Hossain}, \citenamefont {Teweldebrhan},\ and\ \citenamefont
  {Balandin}}]{Shahil2010}%
  \BibitemOpen
  \bibfield  {author} {\bibinfo {author} {\bibfnamefont {K.~M.}\ \bibnamefont
  {Shahil}}, \bibinfo {author} {\bibfnamefont {M.~Z.}\ \bibnamefont {Hossain}},
  \bibinfo {author} {\bibfnamefont {D.}~\bibnamefont {Teweldebrhan}},\ and\
  \bibinfo {author} {\bibfnamefont {A.~A.}\ \bibnamefont {Balandin}},\
  }\bibfield  {title} {\bibinfo {title} {{Crystal symmetry breaking in
  few-quintuple {Bi$_2$Te$_3$ films: Applications in nanometrology of
  topological insulators}}},\ }\href {https://doi.org/10.1063/1.3396190}
  {\bibfield  {journal} {\bibinfo  {journal} {Appl. Phys. Lett.}\ }\textbf
  {\bibinfo {volume} {96}},\ \bibinfo {pages} {153103} (\bibinfo {year}
  {2010})}\BibitemShut {NoStop}%
\bibitem [{\citenamefont {Shahil}\ \emph {et~al.}(2012)\citenamefont {Shahil},
  \citenamefont {Hossain}, \citenamefont {Goyal},\ and\ \citenamefont
  {Balandin}}]{Shahil2012}%
  \BibitemOpen
  \bibfield  {author} {\bibinfo {author} {\bibfnamefont {K.~M.}\ \bibnamefont
  {Shahil}}, \bibinfo {author} {\bibfnamefont {M.~Z.}\ \bibnamefont {Hossain}},
  \bibinfo {author} {\bibfnamefont {V.}~\bibnamefont {Goyal}},\ and\ \bibinfo
  {author} {\bibfnamefont {A.~A.}\ \bibnamefont {Balandin}},\ }\bibfield
  {title} {\bibinfo {title} {{Micro-Raman spectroscopy of mechanically
  exfoliated few-quintuple layers of {Bi$_2$Te$_3$}, {Bi$_2$Se$_3$}, and
  {Sb$_2$Te$_3$} materials}},\ }\href {https://doi.org/10.1063/1.3690913}
  {\bibfield  {journal} {\bibinfo  {journal} {J. Appl. Phys.}\ }\textbf
  {\bibinfo {volume} {111}},\ \bibinfo {pages} {54305} (\bibinfo {year}
  {2012})}\BibitemShut {NoStop}%
\bibitem [{\citenamefont {Richter}\ \emph {et~al.}(1982)\citenamefont
  {Richter}, \citenamefont {Krost}, \citenamefont {Nowak},\ and\ \citenamefont
  {Anastassakis}}]{Richter1982}%
  \BibitemOpen
  \bibfield  {author} {\bibinfo {author} {\bibfnamefont {W.}~\bibnamefont
  {Richter}}, \bibinfo {author} {\bibfnamefont {A.}~\bibnamefont {Krost}},
  \bibinfo {author} {\bibfnamefont {U.}~\bibnamefont {Nowak}},\ and\ \bibinfo
  {author} {\bibfnamefont {E.}~\bibnamefont {Anastassakis}},\ }\bibfield
  {title} {\bibinfo {title} {{Anisotropy and dispersion of coupled
  plasmon-LO-phonon modes in {Sb$_{2}$}{Te$_{3}$}}},\ }\href
  {https://doi.org/10.1007/BF01313026} {\bibfield  {journal} {\bibinfo
  {journal} {Zeitschrift f{\"{u}}r Phys. B Condens. Matter}\ }\textbf {\bibinfo
  {volume} {49}},\ \bibinfo {pages} {191} (\bibinfo {year} {1982})}\BibitemShut
  {NoStop}%
\bibitem [{\citenamefont {Nagao}\ \emph {et~al.}(1998)\citenamefont {Nagao},
  \citenamefont {Hatta},\ and\ \citenamefont {Mukasa}}]{Nagao1998}%
  \BibitemOpen
  \bibfield  {author} {\bibinfo {author} {\bibfnamefont {J.}~\bibnamefont
  {Nagao}}, \bibinfo {author} {\bibfnamefont {E.}~\bibnamefont {Hatta}},\ and\
  \bibinfo {author} {\bibfnamefont {K.}~\bibnamefont {Mukasa}},\ }\bibfield
  {title} {\bibinfo {title} {{Inelastic electron tunneling into
  {p-Bi$_2$Te$_3$} crystals}},\ }\href {https://doi.org/10.1063/1.367274}
  {\bibfield  {journal} {\bibinfo  {journal} {J. Appl. Phys.}\ }\textbf
  {\bibinfo {volume} {83}},\ \bibinfo {pages} {4807} (\bibinfo {year}
  {1998})}\BibitemShut {NoStop}%
\bibitem [{\citenamefont {Richter}\ and\ \citenamefont
  {Becker}(1977)}]{Richter1977}%
  \BibitemOpen
  \bibfield  {author} {\bibinfo {author} {\bibfnamefont {W.}~\bibnamefont
  {Richter}}\ and\ \bibinfo {author} {\bibfnamefont {C.~R.}\ \bibnamefont
  {Becker}},\ }\bibfield  {title} {\bibinfo {title} {{A Raman and
  far‐infrared investigation of phonons in the rhombohedral {V$_2$–VI$_3$}
  compounds {Bi$_2$Te$_3$}, {Bi$_2$Se$_3$}, {Sb$_2$Te$_3$} and
  {Bi$_2$(Te$_{1-x}$Se$_x$)$_3$} (0 {\textless} x {\textless} 1),
  {(Bi$_{1-y}$Sb$_y$)$_2$Te$_3$} (0 {\textless} y {\textless} 1)}},\ }\href
  {https://doi.org/10.1002/pssb.2220840226} {\bibfield  {journal} {\bibinfo
  {journal} {Phys. Status Solidi}\ }\textbf {\bibinfo {volume} {84}},\ \bibinfo
  {pages} {619} (\bibinfo {year} {1977})}\BibitemShut {NoStop}%
\bibitem [{\citenamefont {Rauh}\ \emph {et~al.}(1981)\citenamefont {Rauh},
  \citenamefont {Geick}, \citenamefont {Kohler}, \citenamefont {Nucker},\ and\
  \citenamefont {Lehner}}]{Rauh1981}%
  \BibitemOpen
  \bibfield  {author} {\bibinfo {author} {\bibfnamefont {H.}~\bibnamefont
  {Rauh}}, \bibinfo {author} {\bibfnamefont {R.}~\bibnamefont {Geick}},
  \bibinfo {author} {\bibfnamefont {H.}~\bibnamefont {Kohler}}, \bibinfo
  {author} {\bibfnamefont {N.}~\bibnamefont {Nucker}},\ and\ \bibinfo {author}
  {\bibfnamefont {N.}~\bibnamefont {Lehner}},\ }\bibfield  {title} {\bibinfo
  {title} {{Generalized phonon density of states of the layer compounds
  {Bi$_2$Se$_3$}, {Bi$_2$Te$_3$}, {Sb$_2$Te$_3$} and
  {Bi$_2$(Te$_{0.5}$Se$_{0.5}$)$_3$}, {(Bi$_{0.5}$Sb$_{0.5}$)$_2$Te$_3$}}},\
  }\href {https://doi.org/10.1088/0022-3719/14/20/009} {\bibfield  {journal}
  {\bibinfo  {journal} {J. Phys. C: Solid State Phys.}\ }\textbf {\bibinfo
  {volume} {14}},\ \bibinfo {pages} {2705} (\bibinfo {year}
  {1981})}\BibitemShut {NoStop}%
\bibitem [{\citenamefont {Serrano-S{\'{a}}nchez}\ \emph
  {et~al.}(2017)\citenamefont {Serrano-S{\'{a}}nchez}, \citenamefont
  {Gharsallah}, \citenamefont {Nemes}, \citenamefont {Biskup}, \citenamefont
  {Varela}, \citenamefont {Mart{\'{i}}nez}, \citenamefont
  {Fern{\'{a}}ndez-D{\'{i}}az},\ and\ \citenamefont
  {Alonso}}]{SerranoSanchez2017}%
  \BibitemOpen
  \bibfield  {author} {\bibinfo {author} {\bibfnamefont {F.}~\bibnamefont
  {Serrano-S{\'{a}}nchez}}, \bibinfo {author} {\bibfnamefont {M.}~\bibnamefont
  {Gharsallah}}, \bibinfo {author} {\bibfnamefont {N.~M.}\ \bibnamefont
  {Nemes}}, \bibinfo {author} {\bibfnamefont {N.}~\bibnamefont {Biskup}},
  \bibinfo {author} {\bibfnamefont {M.}~\bibnamefont {Varela}}, \bibinfo
  {author} {\bibfnamefont {J.~L.}\ \bibnamefont {Mart{\'{i}}nez}}, \bibinfo
  {author} {\bibfnamefont {M.~T.}\ \bibnamefont {Fern{\'{a}}ndez-D{\'{i}}az}},\
  and\ \bibinfo {author} {\bibfnamefont {J.~A.}\ \bibnamefont {Alonso}},\
  }\bibfield  {title} {\bibinfo {title} {{Enhanced figure of merit in
  nanostructured {(Bi,Sb)$_{2}$Te$_{3}$} with optimized composition, prepared
  by a straightforward arc-melting procedure}},\ }\href
  {https://doi.org/10.1038/s41598-017-05428-4} {\bibfield  {journal} {\bibinfo
  {journal} {Sci. Rep.}\ }\textbf {\bibinfo {volume} {7}},\ \bibinfo {pages}
  {6277} (\bibinfo {year} {2017})}\BibitemShut {NoStop}%
\bibitem [{\citenamefont {Cheng}\ and\ \citenamefont {Ren}(2011)}]{Cheng2011}%
  \BibitemOpen
  \bibfield  {author} {\bibinfo {author} {\bibfnamefont {W.}~\bibnamefont
  {Cheng}}\ and\ \bibinfo {author} {\bibfnamefont {S.~F.}\ \bibnamefont
  {Ren}},\ }\bibfield  {title} {\bibinfo {title} {{Phonons of single quintuple
  Bi2Te3 and {Bi$_2$Se$_3$} films and bulk materials}},\ }\href
  {https://doi.org/10.1103/PhysRevB.83.094301} {\bibfield  {journal} {\bibinfo
  {journal} {Phys. Rev. B}\ }\textbf {\bibinfo {volume} {83}},\ \bibinfo
  {pages} {94301} (\bibinfo {year} {2011})}\BibitemShut {NoStop}%
\bibitem [{\citenamefont {Hellman}\ and\ \citenamefont
  {Broido}(2014)}]{Hellman2014}%
  \BibitemOpen
  \bibfield  {author} {\bibinfo {author} {\bibfnamefont {O.}~\bibnamefont
  {Hellman}}\ and\ \bibinfo {author} {\bibfnamefont {D.~A.}\ \bibnamefont
  {Broido}},\ }\bibfield  {title} {\bibinfo {title} {{Phonon thermal transport
  in {Bi$_2$Te$_3$} from first principles}},\ }\href
  {https://doi.org/10.1103/PhysRevB.90.134309} {\bibfield  {journal} {\bibinfo
  {journal} {Phys. Rev. B}\ }\textbf {\bibinfo {volume} {90}},\ \bibinfo
  {pages} {134309} (\bibinfo {year} {2014})}\BibitemShut {NoStop}%
\bibitem [{\citenamefont {Voss}(1978)}]{Voss1978}%
  \BibitemOpen
  \bibfield  {author} {\bibinfo {author} {\bibfnamefont {R.~F.}\ \bibnamefont
  {Voss}},\ }\bibfield  {title} {\bibinfo {title} {{1/f noise and percolation
  in impurity bands in inversion layers}},\ }\href
  {https://doi.org/10.1088/0022-3719/11/23/006} {\bibfield  {journal} {\bibinfo
   {journal} {J. Phys. C Solid State Phys.}\ }\textbf {\bibinfo {volume}
  {11}},\ \bibinfo {pages} {L923} (\bibinfo {year} {1978})}\BibitemShut
  {NoStop}%
\bibitem [{\citenamefont {Thouless}(1974)}]{Thouless1974}%
  \BibitemOpen
  \bibfield  {author} {\bibinfo {author} {\bibfnamefont {D.~J.}\ \bibnamefont
  {Thouless}},\ }\bibfield  {title} {\bibinfo {title} {{Electrons in disordered
  systems and the theory of localization}},\ }\href
  {https://doi.org/10.1016/0370-1573(74)90029-5} {\bibfield  {journal}
  {\bibinfo  {journal} {Phys. Rep.}\ }\textbf {\bibinfo {volume} {13}},\
  \bibinfo {pages} {93} (\bibinfo {year} {1974})}\BibitemShut {NoStop}%
\bibitem [{\citenamefont {McWhorter}(1955)}]{McWhorter1955}%
  \BibitemOpen
  \bibfield  {author} {\bibinfo {author} {\bibfnamefont {A.~L.}\ \bibnamefont
  {McWhorter}},\ }\bibfield  {title} {\bibinfo {title} {{PhD. Thesis, Lincoln
  Laboratory}},\ }\href {https://dspace.mit.edu/handle/1721.1/12061} {\bibfield
   {journal} {\bibinfo  {journal} {Technical Report No. 80}\ } (\bibinfo {year}
  {1955})}\BibitemShut {NoStop}%
\bibitem [{\citenamefont {Sah}\ and\ \citenamefont
  {Hielscher}(1966)}]{Sah1966}%
  \BibitemOpen
  \bibfield  {author} {\bibinfo {author} {\bibfnamefont {C.~T.}\ \bibnamefont
  {Sah}}\ and\ \bibinfo {author} {\bibfnamefont {F.~H.}\ \bibnamefont
  {Hielscher}},\ }\bibfield  {title} {\bibinfo {title} {{Evidence of the
  surface origin of the 1/f noise}},\ }\href
  {https://doi.org/10.1103/PhysRevLett.17.956} {\bibfield  {journal} {\bibinfo
  {journal} {Phys. Rev. Lett.}\ }\textbf {\bibinfo {volume} {17}},\ \bibinfo
  {pages} {956} (\bibinfo {year} {1966})}\BibitemShut {NoStop}%
\bibitem [{\citenamefont {Hsu}\ \emph {et~al.}(1968)\citenamefont {Hsu},
  \citenamefont {Fitzgerald},\ and\ \citenamefont {Grove}}]{Hsu1968}%
  \BibitemOpen
  \bibfield  {author} {\bibinfo {author} {\bibfnamefont {S.~T.}\ \bibnamefont
  {Hsu}}, \bibinfo {author} {\bibfnamefont {D.~J.}\ \bibnamefont
  {Fitzgerald}},\ and\ \bibinfo {author} {\bibfnamefont {A.~S.}\ \bibnamefont
  {Grove}},\ }\bibfield  {title} {\bibinfo {title} {{Surface-state related l/f
  noise in p-n junctions and MOS transistors}},\ }\href
  {https://doi.org/10.1063/1.1651995} {\bibfield  {journal} {\bibinfo
  {journal} {Appl. Phys. Lett.}\ }\textbf {\bibinfo {volume} {12}},\ \bibinfo
  {pages} {287} (\bibinfo {year} {1968})}\BibitemShut {NoStop}%
\bibitem [{\citenamefont {Koslowski}\ \emph {et~al.}(1993)\citenamefont
  {Koslowski}, \citenamefont {Baur}, \citenamefont {M{\"{o}}ller},\ and\
  \citenamefont {Dransfeld}}]{Koslowski1993}%
  \BibitemOpen
  \bibfield  {author} {\bibinfo {author} {\bibfnamefont {B.}~\bibnamefont
  {Koslowski}}, \bibinfo {author} {\bibfnamefont {C.}~\bibnamefont {Baur}},
  \bibinfo {author} {\bibfnamefont {R.}~\bibnamefont {M{\"{o}}ller}},\ and\
  \bibinfo {author} {\bibfnamefont {K.}~\bibnamefont {Dransfeld}},\ }\bibfield
  {title} {\bibinfo {title} {{Atomic scale variation of current noise on
  GaAs(110) detected by a scanning tunneling microscope}},\ }\href
  {https://doi.org/10.1016/0039-6028(93)90360-V} {\bibfield  {journal}
  {\bibinfo  {journal} {Surf. Sci.}\ }\textbf {\bibinfo {volume} {280}},\
  \bibinfo {pages} {106} (\bibinfo {year} {1993})}\BibitemShut {NoStop}%
\bibitem [{\citenamefont {Sugita}\ \emph {et~al.}(1996)\citenamefont {Sugita},
  \citenamefont {Mera},\ and\ \citenamefont {Maeda}}]{Sugita1996}%
  \BibitemOpen
  \bibfield  {author} {\bibinfo {author} {\bibfnamefont {S.}~\bibnamefont
  {Sugita}}, \bibinfo {author} {\bibfnamefont {Y.}~\bibnamefont {Mera}},\ and\
  \bibinfo {author} {\bibfnamefont {K.}~\bibnamefont {Maeda}},\ }\bibfield
  {title} {\bibinfo {title} {{Origin of low frequency noise and 1/f
  fluctuations of tunneling current in scanning tunneling microscopes}},\
  }\href {https://doi.org/10.1063/1.361783} {\bibfield  {journal} {\bibinfo
  {journal} {J. Appl. Phys.}\ }\textbf {\bibinfo {volume} {79}},\ \bibinfo
  {pages} {4166} (\bibinfo {year} {1996})}\BibitemShut {NoStop}%
\bibitem [{\citenamefont {Fleetwood}\ \emph {et~al.}(1983)\citenamefont
  {Fleetwood}, \citenamefont {Masden},\ and\ \citenamefont
  {Giordano}}]{Fleetwood1983}%
  \BibitemOpen
  \bibfield  {author} {\bibinfo {author} {\bibfnamefont {D.~M.}\ \bibnamefont
  {Fleetwood}}, \bibinfo {author} {\bibfnamefont {J.~T.}\ \bibnamefont
  {Masden}},\ and\ \bibinfo {author} {\bibfnamefont {N.}~\bibnamefont
  {Giordano}},\ }\bibfield  {title} {\bibinfo {title} {{1/f noise in platinum
  films and ultrathin platinum wires: Evidence for a common, bulk origin}},\
  }\href {https://doi.org/10.1103/PhysRevLett.50.450} {\bibfield  {journal}
  {\bibinfo  {journal} {Phys. Rev. Lett.}\ }\textbf {\bibinfo {volume} {50}},\
  \bibinfo {pages} {450} (\bibinfo {year} {1983})}\BibitemShut {NoStop}%
\bibitem [{\citenamefont {Zimmerman}\ \emph {et~al.}(1986)\citenamefont
  {Zimmerman}, \citenamefont {Scofield}, \citenamefont {Mantese},\ and\
  \citenamefont {Webb}}]{Zimmerman1986}%
  \BibitemOpen
  \bibfield  {author} {\bibinfo {author} {\bibfnamefont {N.~M.}\ \bibnamefont
  {Zimmerman}}, \bibinfo {author} {\bibfnamefont {J.~H.}\ \bibnamefont
  {Scofield}}, \bibinfo {author} {\bibfnamefont {J.~V.}\ \bibnamefont
  {Mantese}},\ and\ \bibinfo {author} {\bibfnamefont {W.~W.}\ \bibnamefont
  {Webb}},\ }\bibfield  {title} {\bibinfo {title} {{Volume versus surface
  origin of 1/f noise in metals}},\ }\href
  {https://doi.org/10.1103/PhysRevB.34.773} {\bibfield  {journal} {\bibinfo
  {journal} {Phys. Rev. B}\ }\textbf {\bibinfo {volume} {34}},\ \bibinfo
  {pages} {773} (\bibinfo {year} {1986})}\BibitemShut {NoStop}%
\bibitem [{\citenamefont {Liu}\ \emph {et~al.}(2013)\citenamefont {Liu},
  \citenamefont {Rumyantsev}, \citenamefont {Shur},\ and\ \citenamefont
  {Balandin}}]{Liu2013}%
  \BibitemOpen
  \bibfield  {author} {\bibinfo {author} {\bibfnamefont {G.}~\bibnamefont
  {Liu}}, \bibinfo {author} {\bibfnamefont {S.}~\bibnamefont {Rumyantsev}},
  \bibinfo {author} {\bibfnamefont {M.~S.}\ \bibnamefont {Shur}},\ and\
  \bibinfo {author} {\bibfnamefont {A.~A.}\ \bibnamefont {Balandin}},\
  }\bibfield  {title} {\bibinfo {title} {{Origin of 1/f noise in graphene
  multilayers: surface vs. volume}},\ }\bibfield  {journal} {\bibinfo
  {journal} {Appl. Phys. Lett.}\ }\textbf {\bibinfo {volume} {102}},\ \href
  {https://doi.org/093111} {093111} (\bibinfo {year} {2013})\BibitemShut
  {NoStop}%
\bibitem [{\citenamefont {Akarvardar}\ \emph {et~al.}(2006)\citenamefont
  {Akarvardar}, \citenamefont {Dufrene}, \citenamefont {Cristoloveanu},
  \citenamefont {Gentil}, \citenamefont {Blalock},\ and\ \citenamefont
  {Mojarradi}}]{Akarvardar2006}%
  \BibitemOpen
  \bibfield  {author} {\bibinfo {author} {\bibfnamefont {K.}~\bibnamefont
  {Akarvardar}}, \bibinfo {author} {\bibfnamefont {B.~M.}\ \bibnamefont
  {Dufrene}}, \bibinfo {author} {\bibfnamefont {S.}~\bibnamefont
  {Cristoloveanu}}, \bibinfo {author} {\bibfnamefont {P.}~\bibnamefont
  {Gentil}}, \bibinfo {author} {\bibfnamefont {B.~J.}\ \bibnamefont
  {Blalock}},\ and\ \bibinfo {author} {\bibfnamefont {M.~M.}\ \bibnamefont
  {Mojarradi}},\ }\bibfield  {title} {\bibinfo {title} {{Low-frequency noise in
  SOI four-gate transistors}},\ }\href
  {https://doi.org/10.1109/TED.2006.870272} {\bibfield  {journal} {\bibinfo
  {journal} {IEEE Trans. Electron Devices}\ }\textbf {\bibinfo {volume} {53}},\
  \bibinfo {pages} {829} (\bibinfo {year} {2006})}\BibitemShut {NoStop}%
\bibitem [{\citenamefont {Mihaila}\ \emph {et~al.}(1991)\citenamefont
  {Mihaila}, \citenamefont {Stepanescu},\ and\ \citenamefont
  {Masoero}}]{Mihaila1991}%
  \BibitemOpen
  \bibfield  {author} {\bibinfo {author} {\bibfnamefont {M.}~\bibnamefont
  {Mihaila}}, \bibinfo {author} {\bibfnamefont {A.}~\bibnamefont
  {Stepanescu}},\ and\ \bibinfo {author} {\bibfnamefont {A.}~\bibnamefont
  {Masoero}},\ }\bibfield  {title} {\bibinfo {title} {{Phonon frequencies in
  the 1/f noise of discontinous platinium films}},\ }in\ \href
  {https://doi.org/10.1063/1.44655} {\emph {\bibinfo {booktitle} {Noise Phys.
  Syst. 1/f Noise}}},\ \bibinfo {series and number} {Ohmsha Ltd},\ \bibinfo
  {editor} {edited by\ \bibinfo {editor} {\bibfnamefont {T.}~\bibnamefont
  {Musha}}, \bibinfo {editor} {\bibfnamefont {S.}~\bibnamefont {Sato}},\ and\
  \bibinfo {editor} {\bibfnamefont {M.}~\bibnamefont {Yamamoto}}}\ (\bibinfo
  {year} {1991})\BibitemShut {NoStop}%
\bibitem [{\citenamefont {Mihaila}(2003)}]{Mihaila2003}%
  \BibitemOpen
  \bibfield  {author} {\bibinfo {author} {\bibfnamefont {M.~N.}\ \bibnamefont
  {Mihaila}},\ }\bibfield  {title} {\bibinfo {title} {{Surface phonons in the
  1/f noise of a discontinuous platinum film}},\ }\href@noop {} {\bibfield
  {journal} {\bibinfo  {journal} {Proc. Rom. Acad. Ser. A,}\ }\textbf {\bibinfo
  {volume} {4}},\ \bibinfo {pages} {229} (\bibinfo {year} {2003})}\BibitemShut
  {NoStop}%
\bibitem [{\citenamefont {Asriyan}\ \emph {et~al.}(2005)\citenamefont {Asriyan}
  \emph {et~al.}}]{Asriyan2005}%
  \BibitemOpen
  \bibfield  {author} {\bibinfo {author} {\bibfnamefont {H.}~\bibnamefont
  {Asriyan}} \emph {et~al.},\ }\bibfield  {title} {\bibinfo {title}
  {{Semiconductor-metal interface influence on the bulk low-frequency noise
  behavior and role of the phonons refraction points}},\ }in\ \href
  {https://doi.org/10.1117/12.609291} {\emph {\bibinfo {booktitle} {Noise Inf.
  Nanoelectron. Sensors, Stand. III}}},\ Vol.\ \bibinfo {volume}
  {$\textbf{5846}$},\ \bibinfo {editor} {edited by\ \bibinfo {editor}
  {\bibfnamefont {J.~A.}\ \bibnamefont {Bergou}}, \bibinfo {editor}
  {\bibfnamefont {J.~M.}\ \bibnamefont {Smulko}}, \bibinfo {editor}
  {\bibfnamefont {M.~I.}\ \bibnamefont {Dykman}},\ and\ \bibinfo {editor}
  {\bibfnamefont {L.}~\bibnamefont {Wang}}}\ (\bibinfo {address} {Austin, TX},\
  \bibinfo {year} {2005})\ p.\ \bibinfo {pages} {192}\BibitemShut {NoStop}%
\bibitem [{\citenamefont {Hammig}\ \emph {et~al.}(2013)\citenamefont {Hammig},
  \citenamefont {Kang}, \citenamefont {Jeong},\ and\ \citenamefont
  {Jarrett}}]{Hammig2013}%
  \BibitemOpen
  \bibfield  {author} {\bibinfo {author} {\bibfnamefont {M.~D.}\ \bibnamefont
  {Hammig}}, \bibinfo {author} {\bibfnamefont {T.}~\bibnamefont {Kang}},
  \bibinfo {author} {\bibfnamefont {M.}~\bibnamefont {Jeong}},\ and\ \bibinfo
  {author} {\bibfnamefont {M.}~\bibnamefont {Jarrett}},\ }\bibfield  {title}
  {\bibinfo {title} {{Suppression of interface-induced noise by the control of
  electron-phonon interactions}},\ }\href
  {https://doi.org/10.1109/TNS.2013.2266798} {\bibfield  {journal} {\bibinfo
  {journal} {IEEE Trans. Nucl. Sci.}\ }\textbf {\bibinfo {volume} {60}},\
  \bibinfo {pages} {2831} (\bibinfo {year} {2013})}\BibitemShut {NoStop}%
\bibitem [{\citenamefont {Toshimitsu}\ \emph {et~al.}(1990)\citenamefont
  {Toshimitsu}, \citenamefont {Borb{\'{e}}ly},\ and\ \citenamefont
  {Shoji}}]{Toshimitsu1990}%
  \BibitemOpen
  \bibfield  {author} {\bibinfo {author} {\bibfnamefont {M.}~\bibnamefont
  {Toshimitsu}}, \bibinfo {author} {\bibfnamefont {G.}~\bibnamefont
  {Borb{\'{e}}ly}},\ and\ \bibinfo {author} {\bibfnamefont {M.}~\bibnamefont
  {Shoji}},\ }\bibfield  {title} {\bibinfo {title} {{1/f phonon-number
  fluctuations in quartz observed by laser light scattering}},\ }\href
  {https://doi.org/10.1103/PhysRevLett.64.2394} {\bibfield  {journal} {\bibinfo
   {journal} {Phys. Rev. Lett.}\ }\textbf {\bibinfo {volume} {64}},\ \bibinfo
  {pages} {2394} (\bibinfo {year} {1990})}\BibitemShut {NoStop}%
\bibitem [{\citenamefont {Voss}\ and\ \citenamefont {Clarke}(1976)}]{Voss1976}%
  \BibitemOpen
  \bibfield  {author} {\bibinfo {author} {\bibfnamefont {R.~F.}\ \bibnamefont
  {Voss}}\ and\ \bibinfo {author} {\bibfnamefont {J.}~\bibnamefont {Clarke}},\
  }\bibfield  {title} {\bibinfo {title} {{1/f noise from systems in thermal
  equilibrium}},\ }\href {https://doi.org/10.1103/PhysRevLett.36.42} {\bibfield
   {journal} {\bibinfo  {journal} {Phys. Rev. Lett.}\ }\textbf {\bibinfo
  {volume} {36}},\ \bibinfo {pages} {42} (\bibinfo {year} {1976})}\BibitemShut
  {NoStop}%
\end{thebibliography}%

\end{document}